\documentclass[onecolumn,nofootinbib]{revtex4}
\usepackage{setspace}
\usepackage{amsmath,amssymb}
\usepackage{graphicx}
\usepackage{dcolumn}
\usepackage{bm,color}
\usepackage{hyperref}
\usepackage{accents}
\usepackage{amssymb,float}
\usepackage{amsmath}
\usepackage{multirow}
\usepackage{tikz}
\usepackage{subcaption}
\usepackage[rightcaption]{sidecap}
\sidecaptionvpos{figure}{t}
\usepackage{enumerate}

\newcolumntype{P}[1]{>{\centering\arraybackslash}p{#1}}
\newcolumntype{M}[1]{>{\centering\arraybackslash}m{#1}}
\hypersetup{
    colorlinks=true,
    linkcolor=red,
    filecolor=magenta,      
    citecolor=blue
}

\newcommand{\udt}[3]{#1^{#2}_{\phantom{#2}#3}}
\newcommand{\udut}[4]{#1^{#2\phantom{#3}#4}_{\phantom{#2}#3\phantom{#4}}}

\newcommand{\dut}[3]{#1_{#2}^{\phantom{#2}#3}}
\newcommand{\dudt}[4]{#1_{#2\phantom{#3}#4}^{\phantom{#2}#3}}
\newcommand{\dd}{\mathrm{d}}
\newcommand{\lc}[1]{\accentset{\circ}{#1}}

\begin{document}

\title{Cosmological Perturbations in the Teleparallel analog of Horndeski gravity}

\author{Bobomurat Ahmedov}
\email{ahmedov@astrin.uz}
\affiliation{Institute of Fundamental and Applied Research,
National Research University TIIAME, Kori Niyoziy 39, Tashkent 100000, Uzbekistan}
\affiliation{National University of Uzbekistan, Tashkent 100174, Uzbekistan}

\author{Konstantinos F. Dialektopoulos}
\email{kdialekt@gmail.com}
\affiliation{Department of Mathematics and Computer Science, Transilvania University of Brasov, 500091, Brasov, Romania}
\affiliation{Laboratory of Physics, Faculty of Engineering, Aristotle University of Thessaloniki, 54124 Thessaloniki, Greece}
\affiliation{Department of Physics, Nazarbayev University, 53 Kabanbay Batyr avenue, 010000 Astana, Kazakhstan}

\author{Jackson Levi Said}
\email{jackson.said@um.edu.mt}
\affiliation{Institute of Space Sciences and Astronomy, University of Malta, Malta, MSD 2080}
\affiliation{Department of Physics, University of Malta, Malta}

\author{Abdurakhmon Nosirov}
\email{abdurahmonnosirov000203@gmail.com}
\affiliation{New Uzbekistan University, Mustaqillik ave. 54, 100007 Tashkent, Uzbekistan}
\affiliation{Ulugh Beg Astronomical Institute, Astronomy St.  33, Tashkent 100052, Uzbekistan}

\author{Zinovia Oikonomopoulou}
\email{zhnobia.oikonomopoulou.21@um.edu.mt}
\affiliation{Institute of Space Sciences and Astronomy, University of Malta, Malta, MSD 2080}

\author{Odil Yunusov}
\email{odilbekhamroev@gmail.com}
\affiliation{New Uzbekistan University, Mustaqillik ave. 54, 100007 Tashkent, Uzbekistan}
\affiliation{Ulugh Beg Astronomical Institute, Astronomy St.  33, Tashkent 100052, Uzbekistan}

\date{\today}

\begin{abstract}
In this work we study the cosmological perturbations in Bahamonde-Dialektopoulos-Levi Said (BDLS) theory, i.e. the teleparallel analog of Horndeski gravity. In order to understand the evolution of structure in a cosmological model, it is necessary to study its cosmology not only in the background but also perturbatively. Both Horndeski and its teleparallel analog have been analyzed a lot in the literature, but in order to study them quantitatively, we need to know their cosmological perturbations. That is why, we study here the scalar-vector-tensor decomposition of the theory and we also express the so-called \textit{alpha} parameters in terms of the arbitrary functions of the theory, that designate the deviation from the $\Lambda$CDM model. We have explored tensor, vector and scalar perturbation of the action up to second order, which drastically opens up new possibilities on searches in the parameter space of scalar-tensor theories in the context of observations.
\end{abstract}

\maketitle

\section{Introduction}\label{sec:intro}

The ever increasing precision in the measurement of the expansion of the Universe \cite{SupernovaSearchTeam:1998fmf,SupernovaCosmologyProject:1998vns} has led to the possibility that the Universe may be expanding faster than predicted by the $\Lambda$CDM concordance model \cite{DiValentino:2021izs}. While the $\Lambda$CDM model has had many theoretical \cite{Peebles:2002gy,Copeland:2006wr,Weinberg:1988cp} and observational open questions \cite{Gaitskell:2004gd}, this would constitute a larger disagreement. Indeed, over the last few years the problem of cosmic tensions has been expressed in several cosmological parameters \cite{DiValentino:2020vhf,DiValentino:2020zio,DiValentino:2020vvd} with the value of the Hubble constant being the parameter most in contention. These cosmic tensions have primarily emerged as divergences between reported cosmological parameter values based either on direct measurements from the late Universe such as Type Ia supernovae, the tip of the red giant branch measurements, strong lensing as well as many other approaches \cite{Riess:2021jrx,Wong:2019kwg,Anderson:2023aga,Freedman:2020dne}, as compared with indirect measurements from the cosmic microwave background (CMB) radiation and big bang nucleosynthesis in addition to other approaches to this regime of the Universe \cite{Aghanim:2018eyx,DES:2021wwk,eBOSS:2020yzd,Zhang:2021yna,Cooke:2017cwo}. The growing tension between cosmological parameters measured either directly or indirectly has prompted a re-evaluation of modifications of the concordance models that have been developed in the literature over the last few decades.

There have been a wide variety of different approaches to modifying the concordance model given the appearance of cosmic tensions in recent years. This has taken a variety of forms such as the reconsideration of the cosmological principle \cite{Krishnan:2021jmh,Krishnan:2021dyb}, early Universe dark energy models \cite{Poulin:2023lkg}, extra degrees of freedom in the form of additional neutrino species in the early Universe \cite{DiValentino:2021imh,DiValentino:2021rjj}, modified gravity \cite{CANTATA:2021ktz} and many others \cite{Addazi:2021xuf}. Modified gravity is particularly interesting since it gives a clear avenue in which to modify the cosmic evolution both at the background and perturbative levels at any regimes in the cosmic history of the Universe. However, there are many directions in which general relativity (GR) can be modified as the gravitational component of the concordance model. One interesting approach that has gained momentum in recent years is metric-affine gravity where the underlying connection on which GR is based is exchanged with other geometries \cite{BeltranJimenez:2019esp,Hehl:1994ue}. This may be a more natural way of modifying GR since it does not require ad hoc conditions on the action. Teleparallel gravity (TG) \cite{Aldrovandi:2013wha} is one such approach in which the curvature associated with the Levi-Civita connection $\lc{\Gamma}^{\sigma}{}_{\mu\nu}$ (over-circles denotes quantities based on the curvature of the Levi-Civita connection in this work) is substituted with a torsional teleparallel connection $\Gamma^{\sigma}{}_{\mu\nu}$.

The teleparallel connection is curvature-free and satisfies metricity \cite{Bahamonde:2021gfp,Krssak:2018ywd,Cai:2015emx}. The result is an altogether novel reformulation of gravitational models which can, following a particular prescription of teleparallel objects produce a teleparallel equivalent of general relativity (TEGR) \cite{Maluf:2013gaa,aldrovandi1995introduction}, which is dynamically equivalent to GR but constructed in a totally different way. Thus, GR and TEGR agree on all classical phenomenology but may be different in the IR limit which may provide more possible directions for quantum theories of gravity. As in curvature based theories of gravity, TEGR can be modified in various different directions. The first modification to TEGR and the most well-known is $f(T)$ gravity \cite{Ferraro:2006jd,Ferraro:2008ey,Bengochea:2008gz,Linder:2010py,Chen:2010va,Bahamonde:2019zea,Paliathanasis:2017htk,Farrugia:2020fcu,Bahamonde:2021srr,Bahamonde:2020bbc} where the TEGR Lagrangian $T$ (the torsion scalar) is generalized to an arbitrary function thereof. This produces a second order theory in terms of metric components and agrees with a growing number of observational phenomena. Another direction in which to modify the TEGR action is to consider it as part of a larger scalar-tensor framework akin to Horndeski gravity \cite{Horndeski:1974wa}.

For curvature based settings, Horndeski gravity is the largest class of second order theories in which a single scalar field is added to the Einstein-Hilbert action. The observation of the gravitational wave event GW170817 \cite{TheLIGOScientific:2017qsa} and its electromagnetic counterpart GRB170817A \cite{Goldstein:2017mmi} has placed severe constraints on this model limiting the most exotic branches of the theory \cite{Ezquiaga:2017ekz}. Within this context, a teleparallel analogue of Horndeski gravity was proposed in Ref.~\cite{Bahamonde:2019shr}. Here, some further conditions had to be placed on the ensuing class of theories since TG tends to observe a wider range of Lovelock terms as compared with curvature based theories~\cite{Gonzalez:2015sha}. This framework of theories has since been further investigated in various scenarios. In Ref.~\cite{Bahamonde:2019ipm} it was found that a much larger class of models are admitted that tolerate a speed of light propagation speed for gravitational waves, while in Ref.~\cite{Bahamonde:2021dqn} the spectrum of polarization modes was analyzed in the context of the various branches of the theory. The post-Newtonian parametrization was studied in Ref.~\cite{Bahamonde:2020cfv} where it was found that most of the exotic ingredients of the theory pass the standard tests in this regime. More recently, the class of models has been explored through the prism of Noether symmetries in Ref.~\cite{Dialektopoulos:2021ryi} where the full classification was developed, while in Refs.~\cite{Bernardo:2021bsg,Bernardo:2021izq} the well-tempering approach developed in Ref.~\cite{Appleby:2018yci} for regular Horndeski gravity was applied in this scenario and where the mechanics of well-tempering was better tuned to the TG setting. There has also been initial work on the stability of the theory with a particular focus on theoretical conditions that can be derived from Minkowski space as described Ref.~\cite{Capozziello:2023foy}.
It is thus critical to analyse in a concrete way the full cosmology of this new teleparallel analogue of Horndeski gravity. The background Friedmann and Klein-Gordon equations can be found in Ref.~\cite{Bahamonde:2019shr}. We are hence motivated to determine the full cosmological perturbation equations around a cosmological background. We do this here together with an initial analysis of these perturbations through some examples. The work is divided as follows, TG and its Horndeski analogue are discussed in Sec.~\ref{sec:Intro_BDLS} while in Sec.~\ref{sec:cosmological_pert} the cosmological perturbations are presented. Some example applications of these perturbation equations are shown in Sec.~\ref{sec:application}. Specifically, we calulcate the tensor and scalar primordial power spectra and we express the tensor-to-scalar ratio in terms of the arbitrary functions of our theory. In addition, we formulate the alpha parameters that could be used to distinguish between $\Lambda$CDM and modified descirptions of cosmology. The main results are then summarized in Sec.~\ref{sec:conc} where we describe the impact of this work within the broader literature on the topic.

\section{Teledeski Gravity: A Teleparallel Analogue to Horndeski Theory} \label{sec:Intro_BDLS}

Let us first provide a brief introduction to teleparallel gravity and its background cosmological dynamics.

\subsection{Teleparallel Gravity} \label{tg_foundations}

Curvature-based theories of gravity, such as GR, depend on a geometric framework in which the Levi-Civita connection $\udt{\lc{\Gamma}}{\sigma}{\mu\nu}$ (over-circles refer to any quantities based on the Levi-Civita connection) is the basis of the geometric objects of the theory, such as the Riemann tensor \cite{misner1973gravitation}. Thus, the Levi-Civita connection is used throughout curvature-based models of gravity such as in the construction of the Einstein-Hilbert action through the Ricci scalar. TG offers a different avenue in which to construct theories of gravity in which the curvature-based connection is replaced with the torsional teleparallel connection $\udt{\Gamma}{\sigma}{\mu\nu}$ \cite{Aldrovandi:2013wha,Bahamonde:2021gfp,Cai:2015emx,Krssak:2018ywd}.

On a more practical level, curvature and torsional theories of gravity differ in that the former is based on the metric tensor $g_{\mu\nu}$ and its derivatives, whereas TG is built using the tetrad $\udt{e}{A}{\mu}$, which accounts for the gravitational variables of the system, and a flat spin connection $\udt{\omega}{B}{C\nu}$. Here, Greek indices refer to coordinates on the general manifold while Latin ones refer to the local Minkowski spacetime. These objects also appear in GR but they are convoluted in that setting making them largely impractical, while in TG the spin connection is an inertial object. The tetrad is directly linked to the metric tensor through
\begin{align}
    g_{\mu\nu} = \udt{e}{A}{\mu}\udt{e}{B}{\nu} \eta_{AB}\quad {\rm and}\quad \eta_{AB} = \dut{E}{A}{\mu}\dut{E}{B}{\nu} g_{\mu\nu}\,,\label{eq:metr_trans}
\end{align}
where $\dut{E}{A}{\mu}$ is the inverse tetrad. Here one can observe that there are an infinite number of possible choices for the tetrad components, and so its the spin connection that acts to retain the diffeomorphism invariance of the system for these different choices. 

The tetrad-spin connection pair represent the possible components for a particular spacetime, and so the teleparallel connection can be written in these terms through \cite{Cai:2015emx,Krssak:2018ywd}
\begin{equation}
    \Gamma^{\lambda}{}_{\nu\mu}=\dut{E}{A}{\lambda}\partial_{\mu}\udt{e}{A}{\nu}+\dut{E}{A}{\lambda}\udt{\omega}{A}{B\mu}\udt{e}{B}{\nu}\,,
\end{equation}
where the spin connection is guaranteed to be flat through the condition \cite{Bahamonde:2021gfp}
\begin{equation}
    \partial_{[\mu}\udt{\omega}{A}{|B|\nu]} + \udt{\omega}{A}{C[\mu}\udt{\omega}{C}{|B|\nu]} \equiv 0\,.
\end{equation}
There also exist unique frames for any spacetime in which the spin connection terms all vanish for particular choices of the tetrad components. This is called the Weitzenb\"{o}ck gauge \cite{Weitzenbock1923}, and is consistently applied when the spin connection field equations identically vanish for these choices.

Gravitational scalars in TG are built by replacing the Levi-Civita connection with its teleparallel analogue. The result of this is that the Riemann tensor vanishes identically $\udt{R}{\alpha}{\beta\gamma\epsilon}(\udt{\Gamma}{\sigma}{\mu\nu}) \equiv 0$ (while the regular Riemann tensor remains nonzero $\udt{\lc{R}}{\alpha}{\beta\gamma\epsilon}(\udt{\lc{\Gamma}}{\sigma}{\mu\nu}) \neq 0$). Thus, we need to define a torsion tensor that is based solely on the teleparallel connection, namely \cite{Aldrovandi:2013wha,ortin2004gravity}
\begin{equation}
    \udt{T}{A}{\mu\nu} := 2\udt{\Gamma}{A}{[\nu\mu]}\,,
\end{equation}
where square brackets denote the antisymmetric operator. The torsion tensor represents the field strength of the theory \cite{Bahamonde:2021gfp}, and is invariant under local Lorentz and diffeomorphic transformations \cite{Krssak:2015oua}.

The torsion tensor can be decomposed into three irreducible parts \cite{PhysRevD.19.3524,Bahamonde:2017wwk}
\begin{align}
    a_{\mu} & :=\frac{1}{6}\epsilon_{\mu\nu\lambda\rho}T^{\nu\lambda\rho}\,,\\[4pt]
    v_{\mu} & :=\udt{T}{\lambda}{\lambda\mu}\,,\\[4pt]
    t_{\lambda\mu\nu} & :=\frac{1}{2}\left(T_{\lambda\mu\nu}+T_{\mu\lambda\nu}\right)+\frac{1}{6}\left(g_{\nu\lambda}v_{\mu}+g_{\nu\mu}v_{\lambda}\right)-\frac{1}{3}g_{\lambda\mu}v_{\nu}\,,
\end{align}
which are the axial, vector, and purely tensorial parts, respectively, and where $\epsilon_{\mu\nu\lambda\rho}$ is the totally antisymmetric Levi-Civita tensor in four dimensions. Using this decomposition, unique gravitational scalar invariants can be built \cite{Bahamonde:2015zma}
\begin{align}
    T_{\text{ax}} & := a_{\mu}a^{\mu} = -\frac{1}{18}\left(T_{\lambda\mu\nu}T^{\lambda\mu\nu}-2T_{\lambda\mu\nu}T^{\mu\lambda\nu}\right)\,,\\[4pt]
    T_{\text{vec}} & :=v_{\mu}v^{\mu}=\udt{T}{\lambda}{\lambda\mu}\dut{T}{\rho}{\rho\mu}\,,\\[4pt]
    T{_{\text{ten}}} & :=t_{\lambda\mu\nu}t^{\lambda\mu\nu}=\frac{1}{2}\left(T_{\lambda\mu\nu}T^{\lambda\mu\nu}+T_{\lambda\mu\nu}T^{\mu\lambda\nu}\right)-\frac{1}{2}\udt{T}{\lambda}{\lambda\mu}\dut{T}{\rho}{\rho\mu}\,,
\end{align}
which form the set of all general scalar invariants that are not parity violating and involve, at most, quadratic contractions of the torsion tensor.

There is a particular combination of the axial, vector, and purely tensorial scalar invariants that produce the so-called torsion scalar \cite{Bahamonde:2021gfp}
\begin{equation}
    T:=\frac{3}{2}T_{\text{ax}}+\frac{2}{3}T_{\text{ten}}-\frac{2}{3}T{_{\text{vec}}}=\frac{1}{2}\left(E_{A}{}^{\lambda}g^{\rho\mu}E_{B}{}^{\nu}+2E_{B}{}^{\rho}g^{\lambda\mu}E_{A}{}^{\nu}+\frac{1}{2}\eta_{AB}g^{\mu\rho}g^{\nu\lambda}\right)T^{A}{}_{\mu\nu}T^{B}{}_{\rho\lambda}\,,
\end{equation}
which turns out to be an incredibly important scalar since it is equal to the Ricci scalar (up to a total divergence term) \cite{Bahamonde:2015zma}
\begin{equation}
    R=\lc{R}+T-\frac{2}{e}\partial_{\mu}\left(e\udut{T}{\lambda}{\lambda}{\mu}\right)=0\,,
\end{equation}
where the torsion connection calculated Ricci scalar $R$ vanishes as described above, and where $e=\text{det}\left(\udt{e}{A}{\mu}\right)=\sqrt{-g}$ is the tetrad determinant. The regular curvature-based Ricci scalar can be equivalently expressed as
\begin{equation}
    \lc{R}=-T+\frac{2}{e}\partial_{\mu}\left(e\udut{T}{\lambda}{\lambda}{\mu}\right) := -T+B\,,
\end{equation}
where $B$ is this total divergence term.

The action that is based on the linear form of the torsion scalar is guaranteed to produce a teleparallel equivalent of general relativity (TEGR) (up to a boundary term), which will be dynamically equivalent \cite{Hehl:1994ue,Aldrovandi:2013wha}. Using the same rationale as in curvature-based theories of gravity, one can direct generalize the TEGR action to an f(T) gravity \cite{Ferraro:2006jd,Ferraro:2008ey,Bengochea:2008gz,Finch:2018gkh,Paliathanasis:2017htk,Linder:2010py,Chen:2010va,Bahamonde:2019zea,Cai:2015emx,Farrugia:2016qqe,Finch:2018gkh,Farrugia:2016xcw,Iorio:2012cm,Deng:2018ncg} in which the Lagrangian of the theory is raised from $T$ to $f(T)$. The f(T) field equations have an advantage over their curvature-based analogue in that they remain generically second order in terms of the metric derivatives at the level of the field equations.

In this work we investigate the teleparallel analogue of Horndeski gravity which is built on the interactions of gravitational objects with a single scalar field. In TG, the scalar fields couple to matter identically as in GR (using the minimal coupling prescription), where partial derivatives are raised to the Levi-Civita covariant derivative, that is \cite{Aldrovandi:2013wha,BeltranJimenez:2020sih}
\begin{equation}
    \partial_{\mu} \rightarrow \mathring{\nabla}_{\mu}\,,
\end{equation}
which only applies to the matter sector. With this in hand, we can review the recently proposed teleparallel analog of Horndeski gravity \cite{Bahamonde:2019shr,Bahamonde:2019ipm,Bahamonde:2020cfv}, also called Bahamonde-Dialektopoulos-Levi Said (BDLS) theory. This is based on three base conditions, namely (i) the field equations must be at most second order in their derivatives of the tetrads; (ii) the scalar invariants will not be parity violating; and (iii) the number of contractions with the torsion tensor is limited to being at most quadratic. Without these conditions, the ensuing action would not be finite in nature. Due, in part, to the second order nature of many extensions of TG, it turns out that the resulting action is an extension of the regular form of Horndeski gravity. The result of these conditions is that the terms of regular Horndeski gravity are found, as well as additional terms that are linear in contractions with the torsion tensor \cite{Bahamonde:2019shr}
\begin{equation}
    I_2 = v^{\mu} \phi_{;\mu}\,,
\end{equation}
where $\phi$ is the scalar field, as well as terms that are quadratic in this respect
\begin{align}
    J_{1} & =a^{\mu}a^{\nu}\phi_{;\mu}\phi_{;\nu}\,,\\[4pt]
    J_{3} & =v_{\sigma}t^{\sigma\mu\nu}\phi_{;\mu}\phi_{;\nu}\,,\\[4pt]
    J_{5} & =t^{\sigma\mu\nu}\dudt{t}{\sigma}{\alpha}{\nu}\phi_{;\mu}\phi_{;\alpha}\,,\\[4pt]
    J_{6} & =t^{\sigma\mu\nu}\dut{t}{\sigma}{\alpha\beta}\phi_{;\mu}\phi_{;\nu}\phi_{;\alpha}\phi_{;\beta}\,,\\[4pt]
    J_{8} & =t^{\sigma\mu\nu}\dut{t}{\sigma\mu}{\alpha}\phi_{;\nu}\phi_{;\alpha}\,,\\[4pt]
    J_{10} & =\udt{\epsilon}{\mu}{\nu\sigma\rho}a^{\nu}t^{\alpha\rho\sigma}\phi_{;\mu}\phi_{;\alpha}\,,
\end{align}
where semicolons represent covariant derivatives with respect to the Levi-Civita connection.

Therefore, we can write the teleparallel analogue of Horndeski gravity as an action through
\begin{equation}\label{action}
    \mathcal{S}_{\text{BDLS}} = \int d^4 x\, e\mathcal{L}_{\text{Tele}} + \sum_{i=2}^{5} \int d^4 x\, e\mathcal{L}_i+ \int d^4x \, e\mathcal{L}_{\rm m}\,,
\end{equation}
where the contributions from regular Horndeski gravity continue to appear as \cite{Horndeski:1974wa}
\begin{align}
    \mathcal{L}_{2} & :=G_{2}(\phi,X)\,,\label{eq:LagrHorn1}\\[4pt]
    \mathcal{L}_{3} & :=-G_{3}(\phi,X)\mathring{\Box}\phi\,,\\[4pt]
    \mathcal{L}_{4} & :=G_{4}(\phi,X)\left(-T+B\right)+G_{4,X}(\phi,X)\left[\left(\mathring{\Box}\phi\right)^{2}-\phi_{;\mu\nu}\phi^{;\mu\nu}\right]\,,\\[4pt]
    \mathcal{L}_{5} & :=G_{5}(\phi,X)\mathring{G}_{\mu\nu}\phi^{;\mu\nu}-\frac{1}{6}G_{5,X}(\phi,X)\left[\left(\mathring{\Box}\phi\right)^{3}+2\dut{\phi}{;\mu}{\nu}\dut{\phi}{;\nu}{\alpha}\dut{\phi}{;\alpha}{\mu}-3\phi_{;\mu\nu}\phi^{;\mu\nu}\,\mathring{\Box}\phi\right]\,,\label{eq:LagrHorn5}
\end{align}
which turn out to be equivalent to their corresponding regular Horndeski contributions except that they are calculated using teleparallel objects rather than the metric, but which continue to produce the same contributions to equations of motion for particular systems, and where
\begin{equation}
\label{eq:LTele}
    \mathcal{L}_{\text{Tele}}:= G_{\text{Tele}}\left(\phi,X,T,T_{\text{ax}},T_{\text{vec}},I_2,J_1,J_3,J_5,J_6,J_8,J_{10}\right)\,,
\end{equation}
where the kinetic term is defined as $X:=-\frac{1}{2}\partial^{\mu}\phi\partial_{\mu}\phi$, $\mathcal{L}_{\rm m}$ is the matter Lagrangian in the Jordan conformal frame, $\lc{G}_{\mu\nu}$ is the standard Einstein tensor, and where commas represent regular partial derivatives. For the limit where $G_{\text{Tele}} = 0$, we recover regular Horndeski gravity.

\subsection{Background Cosmology} \label{sec:TG_FLRW}

By varying the action with respect to the tetrad, spin connection and scalar field, we can determine the field equations, as presented in Ref.~\cite{Bahamonde:2020cfv}. The increased generality of the theory produces vastly larger field equations. For this reason, we consider only the equations of motion for a (maximally symmetric) flat FLRW cosmology
\begin{equation}
    \dd s^2 = -N(t)^2 \dd t^2 + a(t)^2(\dd x^2 + \dd y^2 + \dd z^2)\,,
\end{equation}
where $N(t)$ is the lapse function (set to unity once the equations of motion are determined), and $a(t)$ is the scale factor. To obtain the modified equations of motion, we consider the tetrad choice $\udt{e}{a}{\mu} = \textrm{diag}(N(t),a(t),a(t),a(t))$ which is compatible with the Weitzenb\"{o}ck gauge \cite{Krssak:2018ywd,Bahamonde:2021gfp}.

Taking a variation with respect to the dynamical variables of $N(t)$, $a(t)$, and $\phi(t)$, we obtain the equations of motion of the system for a flat homogeneous and isotropic background. This results in the Friedmann equation
\begin{equation}
    \mathcal{E}_{\rm Tele} + \sum_{i=2}^5 \mathcal{E}_i = 0\,,
\end{equation}
where
\begin{align}
    \mathcal{E}_{\rm Tele} &= 6 H\dot{\phi}\tilde{G}_{6,I_2}+12 H^2 \tilde{G}_{6,T}+2X \tilde{G}_{6,X}-\tilde{G}_{6}\,,\\
    \mathcal{E}_2 &= 2XG_{2,X}-G_2\,,\\
    \mathcal{E}_3 &= 6X\dot\phi HG_{3,X}-2XG_{3,\phi}\,,\\
    \mathcal{E}_4 &= -6H^2G_4+24H^2X(G_{4,X}+XG_{4,XX}) - 12HX\dot\phi G_{4,\phi X}-6H\dot\phi G_{4,\phi }\,,\\
    \mathcal{E}_5 &= 2H^3X\dot\phi\left(5G_{5,X}+2XG_{5,XX}\right) - 6H^2X\left(3G_{5,\phi}+2XG_{5,\phi X}\right)\,,
\end{align}
and
\begin{equation}
    \mathcal{L}_{\rm Tele}=\tilde{G}_6(\phi,X,T,I_{2})\,,
\end{equation}
which represents all the nonvanishing scalars for $G_{\rm{Tele}}$, the Hubble parameter is defined as $H = \dot{a}/a$, and dots denote derivatives with respect to cosmic time. The torsion scalar takes on the form $T = 6H^2$, while $I_2 = 3H\dot{\phi}$ and $X = \frac{1}{2} \dot{\phi}^2$, and commas denote partial derivatives. Taking a variation with respect to the scale factor leads to the second Friedmann equation
\begin{equation}
    \mathcal{P}_{\rm Tele}+\sum_{i=2}^5 \mathcal{P}_i=0\,,
\end{equation}
where
\begin{align}
    \mathcal{P}_{\rm Tele}&=-3 H\dot{\phi}\tilde{G}_{6,I_2}-12 H^2\tilde{G}_{6,T}-\frac{d}{dt}\Big(4H \tilde{G}_{6,T}+\dot{\phi}\,\tilde{G}_{6,I_2}\Big)+\tilde{G}_6\,,\\
    \mathcal{P}_2&=G_2\,,\\
    \mathcal{P}_3&=-2X\left(G_{3,\phi}+\ddot\phi G_{3,X} \right) \,,\\
    \mathcal{P}_4&=2\left(3H^2+2\dot H\right) G_4 - 12 H^2 XG_{4,X}-4H\dot X G_{4,X} - 8\dot HXG_{4,X}\nonumber\\
    & \phantom{gggg}-8HX\dot X G_{4,XX} +2\left(\ddot\phi+2H\dot\phi\right) G_{4,\phi} + 4XG_{4,\phi\phi} + 4X\left(\ddot\phi-2H\dot\phi\right) G_{4,\phi X}\,,\\
    \mathcal{P}_5&=-2X\left(2H^3\dot\phi+2H\dot H\dot\phi+3H^2\ddot\phi\right)G_{5,X} - 4H^2X^2\ddot\phi G_{5,XX}\nonumber\\
    & \phantom{gggg} +4HX\left(\dot X-HX\right)G_{5,\phi X} + 2\left[2\frac{d}{dt}\left(HX\right)+3H^2X\right]G_{5,\phi} + 4HX\dot\phi G_{5,\phi\phi}\,.
\end{align}
Finally, the modified Klein-Gordon equation can be determined by taking a variation with respect to the scalar field, giving
\begin{equation}
    \frac{1}{a^3}\frac{\dd}{\dd t}\Big[a^3 (J+J_{\rm Tele})\Big]=P_{\phi}+P_{\rm Tele}\,,
\end{equation}
where the standard Horndeski terms appear as $J$ and $P_{\phi}$ which come from the Lagrangian terms $\mathcal{L}_i$, where $i=2,..,5$ and \cite{Kobayashi:2011nu}
\begin{align}
    J &= \dot\phi G_{2,X} +6HXG_{3,X}-2\dot\phi G_{3,\phi} + 6H^2\dot\phi\left(G_{4,X}+2XG_{4,XX}\right)-12HXG_{4,\phi X}\nonumber\\
    & \phantom{gggggggg} + 2H^3 X\left(3G_{5,X}+2XG_{5,XX}\right) - 6H^2\dot\phi\left(G_{5,\phi}+XG_{5,\phi X}\right)\,,\\
    P_{\phi} &= G_{2,\phi} -2X\left(G_{3,\phi\phi}+\ddot\phi G_{3,\phi X}\right) + 6\left(2H^2+\dot H\right)G_{4,\phi} \nonumber\\
    & \phantom{gggggii} + 6H\left(\dot X+2HX\right)G_{4,\phi X} -6H^2XG_{5,\phi\phi}+2H^3X\dot\phi G_{5,\phi X}\,,
\end{align}
while $J_{\rm Tele}$ and $P_{\rm Tele}$ are new terms related to the teleparallel Horndeski, given by
\begin{align}
    J_{\rm Tele} &= \dot{\phi}\tilde{G}_{6,X}\,,\\
    P_{\rm Tele} &= -9 H^2\tilde{G}_{6,I_2}+\tilde{G}_{6,\phi}-3  \frac{\dd}{\dd t}\left(H\tilde{G}_{6,I_2}\right)\,.
\end{align}
Interestingly, the arguments of $\tilde{G}_6$ do not depend on $T_{\rm ax}$ and $T_{\rm ten}$ since they are zero for the flat FLRW metric. For this reason, we can simply write the contributions in  terms of $T_{\rm vec}$ since $T = -(2/3)T_{\rm vec} = 6H^2$ for this case. Naturally, one should not expect this to also be true at perturbative level \cite{Bahamonde:2019ipm}. Indeed, in this work. explore the cosmological perturbations of the teleparallel analogue of Horndeski gravity. This will open the way for deeper investigations into specific models of the framework that may lead to a better understanding of which models are most compatible with observations of the cosmic evolution of the Universe.

\section{Cosmological Perturbations}\label{sec:cosmological_pert}

The goal of cosmological perturbations in general, is to associate the physics of the early Universe to CMB anisotropies and large-scale structure and to provide the initial conditions for numerical simulations of structure formation.
In this scenery, the physical quantities can be decomposed into a homogeneous background where their dependence is restricted only to the cosmic time and the perturbative parts which depend on spacetime. Linear perturbations around a spatially flat FLRW spacetime in Horndeski gravity have been studied in \cite{Kobayashi:2011nu}. Following that, our starting point is to establish the adequate form of the tetrad matrix and then proceed to the evaluation of the action for tensor, vector and scalar perturbations. In what follows, we work on the unitary gauge, where the scalar field perturbations vanish, i.e. $\delta \phi = 0$.

The first-order expansion of the metric can be parametrized as
\begin{equation}
     g_{\mu\nu} = g_{\mu\nu}^0 + \delta g_{\mu\nu}\,,
\end{equation}
where $g_{\mu\nu}^0 $ in our case is the spatially flat FLRW background, while $\delta g_{\mu\nu}$ can be expressed in an irreducible form as

\begin{align} \label{metric perturbation}
     \delta  g _{\mu\nu} \rightarrow  \left(
\begin{matrix}
    -2\alpha & a B _i +a^2 \partial _i \beta \\
     a B _i + a^2 \partial _i \beta  & 2 a ^2 \left[ \zeta \delta _{ij}+ \frac{1}{2} h _{ij} +2\partial_{(i} 
 h _{j)}\right]
\end{matrix}
\right),
\end{align}
where $h_{ij}$ is symmetric, traceless ($h_{ij} \delta^{ij}$) and divergencless ($\partial^i h_{ij}$) tensor perturbation, $B_i=b_i+\beta_i$, $h_i$ are solinoidal vector perturbations (i.e. $\partial_i X^i=0$) and $\alpha$, $\beta$ as well as $\zeta$ are scalar perturbations. According to \eqref{eq:metr_trans} there are many tetrads that could reproduce the above metric. In what follows, we choose to work with 
\begin{equation}
    e^A{}_{\mu} = \bar{e}^A{}_{\mu} + \delta e^A{}_{\mu}\,,
    \label{fulltetrad}
\end{equation}
where $\bar{e}^A{}_\mu = {\rm diag} (N, a, a, a)$ is the background flat FLRW tetrad and $\delta e^A{}_{\mu}$ is the perturbed tetrad that is given by

\begin{align}
    \delta e ^A {}_\mu \rightarrow  \left(
\begin{matrix}
    \alpha & -a \beta _{i} \\
     \delta ^I _i( \partial ^i \beta + b ^i) & a \delta ^{Ii}\left[ \zeta \delta _{ij}+  \frac{1}{2} h _{ij} + \frac{1}{8} h _{ik} h _{kj} +2 \partial _{(i} 
 h _{j)}\right]
\end{matrix}
\right),
\label{tetrad}
\end{align}

where capital Latin indices indicate components at the Lorentz spacetime, while small ones, the associated 3 dimensional spatial part. For greater details one can check \cite{Bahamonde:2021gfp,Capozziello:2023foy}.

\subsection{Tensor perturbation}\label{subsec:tensor_pert}

We consider the tensor perturbations of Eq.~(\ref{tetrad}) to be described by two functions of spacetime, $h_+(t,x,y,z)$ and $h_\times(t,x,y,z)$ as the two components of a symmetric, traceless and divergenceless tensor $h_{ij}$. We assume for the sake of simplicity, that the perturbations lie on the x-y plane, so that the z-axis is in the direction of the wavevector $\vec{k}$, meaning $\hat{k} = z$. Setting $a, \beta, \zeta$ and $b_i, \beta_i, h_i$ to zero, the tetrad (\ref{fulltetrad}) takes the form
\begin{align}
\udt{e}{A}{\mu} \rightarrow \left( 
\begin{matrix}
1&0&0&0\\
0& -a - \frac{1}{2}  a h_+ -\frac{1}{8} a \left(h_+ ^2 + h_\times ^2\right) & -\frac{1}{2} a h_\times & 0\\
0&-\frac{1}{2} a h_\times &-a +\frac{1}{2} a h_+ -\frac{1}{8} a \left(h_+^2+h_\times ^2\right),&0\\
0&0&0&a
\end{matrix}
\right),
\end{align}
where the $a$ stands for the scale factor. The associated metric from the Eq.~\eqref{eq:metr_trans} becomes
\begin{equation}
g_{\mu\nu} = -N^2 \dd t^2 + a^2 (\delta _{ij} + h_{ij})\dd x^i \dd x ^j\,,
\end{equation}
where the background value of the lapse function is considered to be unity. The scalar invariants of the decomposition of the torsion tensor become 
\begin{align}
    T_{\rm ax}&=\,\,\,\,\,0\ ,\\[4pt]
    T_{\rm vec}&=\,-\frac{9\dot{a}^2}{a^2}\ ,\\[4pt]
    T_{\rm ten}&=\,\,\,\frac{3}{4a^2}\,\left[\,(\bm{\nabla}h_{+})^2+(\bm{\nabla}h_{\times})^2-a^2\dot{h}_{+}^2-a^2\dot{h}_{\times}^2\,\right]\, .
\end{align}

Taking these into account, the action \eqref{action} becomes up to second-order perturbations,
\begin{equation}
    S_{\rm T}^{(2)} = \int \dd t \dd ^3 x \, \,\frac{a^3}{4} \left[\,\mathcal{G}_T \dot{h}_{ij}^2 - \frac{\mathcal{F}_T}{a^2}(\bm{\nabla} h_{ij})^2\,\right],\label{eq:ten_pert_action}
\end{equation}
where
\begin{gather}
    \mathcal{G}_{\rm T} = 2 \left(G_4-2 X G_{4,X}+ X G_{5,\phi}-H X \dot{\phi} G_{5,X} + 2 X G_{{\rm Tele}, J_8}+\frac{X}{2} G_{{\rm Tele}, J_5} -  G_{{\rm Tele},T}\right),
    \label{scalarcoef1} \\
    \mathcal{F}_{\rm T} = 2 \left(G_4-X G_{5,\phi}-X \ddot{\phi} G_{5,X}-G_{{\rm Tele},T} \right),
    \label{scalarcoef2}
\end{gather}
where $G_{i,A}$ denotes the derivative of $G_i$ with respect to $A$. Notice that in the limiting case, where $G_{\rm Tele}$ vanishes, Eq.~\eqref{scalarcoef1} and \eqref{scalarcoef2}, the standard Hordeski results are recovered for the perturbed action~\cite{Kobayashi:2011nu}.

This action represents how tensor perturbations propagate on a flat FLRW background cosmology. By taking a variation of Eq.~\eqref{eq:ten_pert_action}, we find the propagation equation of gravitational waves, which confirms the result in Ref.~\cite{Bahamonde:2019ipm}. We can similarly use this result in the standard $\alpha_i$ parametrization of cosmological perturbations, as we do later on. Specifically, the squared speed of the gravitational waves is given by
\begin{equation}
    c_{\rm T} ^2 = \frac{\mathcal{F}_{\rm T}}{\mathcal{G}_{\rm T}}\,,
\end{equation}
which is not equal to unity for arbitrary $G_i$'s. In addition, one can see from the action \eqref{eq:ten_pert_action} that the following conditions should hold 
\begin{equation}
    \mathcal{F}_{\rm T} > 0 \quad \text{and} \quad \mathcal{G}_{\rm T} > 0 \, ,
    \label{tensorstability}
\end{equation}
in order to avoid ghost and gradient instabilities.

\subsection{Vector perturbation}
\label{subsec:vector_pert}

Another important component of the cosmological perturbations decomposition is vector perturbations, which are divergence-free in nature. Usually, vector perturbations decay in an expanding background cosmology, unless they are driven by anisotropic stress. We do not expect here to happen something different, but to be sure we have to study them in detail. Fixing $\alpha$, $\beta$, $\zeta$ and $h_{ij}$ to vanish in (\ref{tetrad}) the tetrad becomes
\begin{align}
    e^A\text{}_\mu \rightarrow  \left(
\begin{matrix}
    1 & -a \beta_{i} \\
    \delta^I_i b^i& a  \delta^{Ii}\left[\delta_{ij}+2\partial_{(i} 
 h_{j)}\right] 
\end{matrix}
\right).
\end{align}
The scalar invariants of the theory (\ref{tetrad}) take the form
\begin{align}
T_{\rm ax}=&\frac{1}{9a^2} \left[\bm{\nabla} \times (\bm{\beta}-\bm{b})\right]^2,
\\ 
T_{\rm vec}=&-\frac{9 \dot{a}^2}{a^2}+\frac{1}{a^2} \Bigg\{ 9 \dot{a}^2 \left(2 \bm{b}\bm{\beta}+\bm{\beta}^2\right)-6 \dot{a} \left[ a \left[\bm{b} \dot{\bm{\beta}}-2 (\bm{\nabla} \times \bm{h})(\bm{\nabla}\times \dot{\bm{h}} )\right]+(\bm{\nabla} \times \bm{b}) (\bm{\nabla} \times \bm{h}) \right]- \\ \nonumber
&-6 \dot{a} (\bm{b}+\bm{\beta}) \left(\bm{\nabla}^2 \bm{h}-a \dot{\bm{\beta}} \right) +\left(\bm{\nabla}^2 \bm{h}-a \dot{\bm{\beta}} \right)^2 \Bigg\},
\\ 
T_{\rm ten}=&-\frac{1}{a^2}\left[a^2 \dot{\bm{\beta}}^2-(\bm{\nabla} \times \bm{b})^2-(\bm{\nabla} \times \bm{b}) (\bm{\nabla} \times \bm{\beta})-(\bm{\nabla}\times \bm{\beta})^2+3a(\bm{\nabla}\times\dot{\bm{h}})(\bm{b}-a \bm{h})+a\dot{\bm{\beta}}\bm{\nabla}^2 \bm{h}+(\nabla^2 \bm{h})^2\right]\ .
\end{align}

%
We expand the action keeping up to quadratic order terms in the perturbations and we get 
\begin{align}
    S^{(2)}_{\rm V}= &\int \dd t \dd^3 x a^3 \Bigg[ \frac{A_1}{a^2}(\bm{\nabla}^2 \bm{h})^2+A_2\dot{\bm{\beta}}^2+ A_3 (\bm{\nabla}\times{\dot{\bm{h}}})^2-\frac{A_3}{a}(\bm{\nabla}\times{\dot{\bm{h}}})(\bm{\nabla}\times{\bm{b}})+\frac{A_4}{a^2} (\bm{\nabla}\times{\bm{b}})^2+\frac{A_5}{a^2} (\bm{\nabla}\times{\bm{\beta}})^2 \\ \nonumber 
    &+ \frac{A_6}{a} (\bm{\nabla}\times{\bm{h}})(\bm{\nabla}\times{\bm{\beta}})+ \frac{A_7}{a}(\bm{\nabla}\times{\bm{h}})(\bm{\nabla}\times{\dot{\bm{\beta}}})+ \frac{A_8}{a^2}(\bm{\nabla}\times{\bm{\beta}})(\bm{\nabla}\times{\bm{b}})\Bigg]\ ,
\end{align}
where
\begin{align}
    A_1:=&\frac{X}{18} \left(2 X G_{{\rm Tele},J_{6}}-6 G_{{\rm Tele},J_3}-5 G_{{\rm Tele},J_5}-2 G_{{\rm Tele},J_8}\right)+ G_{{\rm Tele},\rm T_{\rm vec}} ,    
\\
     A_2:=&\frac{2 X}{9}\left(2 X G_{{\rm Tele},J_6}+3 G_{{\rm Tele},J_3}-5 G_{{\rm Tele},J_5}-2 G_{{\rm Tele},J_8}\right)+G_{{\rm Tele},T_{\rm vec}},
\\
    A_3:=& X \left(-4 G_{4,X}-2 G_{5,X} H \dot{\phi}+2 G_{5,\phi }+G_{{\rm Tele},J_5}+4 G_{{\rm Tele},J_8}\right)+2 \left(G_{4}-G_{{\rm Tele},T}\right) ,
\\
    A_4:=& \frac{a}{18}\left[3 X \left(-6 G_{4X}-3 G_{5X} H \dot{\phi}+3 G_{5\phi}+3 G_{\rm Tele,J5}+6 G_{\rm Tele,J8}+2 G_{\rm Tele,J10}\right)+9 G_{4}-9 G_{\rm Tele,T}+2 G_{\rm Tele,T_{ax}}\right]\ ,
\\
    A_5:=& \frac{X}{2} \left(-2 G_{4,X}-G_{5,X} H \dot{\phi}+G_{5,\phi}+2 G_{{\rm Tele},J_5}\right)+\frac{1}{2}G_{4}-\frac{1}{2}G_{{\rm Tele},T}+\frac{1}{9}G_{{\rm Tele},T_{\rm ax}}-\frac{2X}{3} G_{{\rm Tele},J_{10}} ,
\\
    A_6:=&-2\frac{d}{dt}(2 X G_{4X}-X G_{5\phi}-G_4)-2\dot{\phi} X G_{5,X} \dot{H}-\dot{\phi}G_{{\rm Tele},I_2}-4 X G_{5.X} H^2 \dot{\phi} \\ \nonumber
    &-H \Bigg\{2 X \left[4 G_{4,X}+\left(2 X G_{5,XX}+3 G_{5,X}\right)\ddot{\phi}+2 X G_{5,\phi X}-2 G_{5,\phi}\right]-4 G_{4}+4 G_{{\rm Tele},T}-6 G_{{\rm Tele},T_{\rm vec}}\Bigg\} ,
\\
    A_7:=&\frac{X}{9} \left(-36 G_{4,X}-18 G_{5,X} H\dot{\phi}+18 G_{5,\phi}-4 X G_{{\rm Tele},J_6}+3 G_{{\rm Tele},J_3}+10 G_{{\rm Tele},J_5}+4 G_{{\rm Tele},J_8}\right)+ \\ \nonumber
    &+2 \left(G_{4}-G_{{\rm Tele},T}+G_{{\rm Tele},T_{\rm vec}}\right) ,
\\
    A_8:=& X \left(-2 G_{4,X}-G_{5,X} H \dot{\phi} +G_{5,\phi}+G_{{\rm Tele},J_5}\right)+G_{4}-G_{{\rm Tele},T}-\frac{2}{9} G_{{\rm Tele},T_{\rm ax}}+\frac{X}{3}G_{{\rm Tele},J_{10}} .
\end{align}

The coefficients of $\bm{\beta}^2, \bm{\beta} \bm{b}$ and $\bm{\nabla} \times \bm{h}$ vanish thanks to background equations. If we choose the condition with $A_1=0$ and $A_3=0$ we have one dynamic variable $\bm{\beta}$ and two auxiliaries fields $\bm{h}$ and $\bm{b}$. By varying with respect to these two auxiliaries fields, we have the following constraint equations:

\begin{gather}
    2\frac{A_4}{a^2} \left(\bm{\nabla} \times \bm{b} \right)+\frac{A_8}{a^2} \left(\bm{\nabla} \times \bm{\beta}\right)=0 ,\\
    \frac{A_6}{a} \left(\bm{\nabla} \times \bm{\beta} \right)+\frac{A_7}{a}\left(\bm{\nabla} \times \dot{\bm{\beta}} \right)=0 .
    \label{constraint1}
\end{gather}
Using Eq.~(\ref{constraint1}) we rewrite the equation for the action as given below:
%
\begin{align}\label{vector}
     S_{\mathrm{V}}^{(2)}&=\int \dd t \dd^3 x a^3\Bigg[\mathcal{G}_V \dot{\bm{\beta}}^2-\frac{\mathcal{F}_V}{a^2}(\bm{\nabla} \times \bm{\beta})^2\Bigg] ,
\end{align}
where
\begin{gather}
    \mathcal{G}_{\rm V} = A_2  \quad \rm{and} \quad \mathcal{F}_{\rm V} = \frac{A_8^2}{4 A_4}-A_5 .
\end{gather}
From Eq.~\eqref{vector} we can see that in order to avoid ghost and gradient instabilities, following inequalities should be satisfied

\begin{align}
    \mathcal{F}_{\rm V}>0, \quad {\rm }\quad \mathcal{G}_{\rm V}>0 .
\end{align}

As in the curvature-based Horndeski cosmology case, these perturbations are found to contribute little to the resulting cosmology, and by and large, do not impact the evolution of the Universe after inflation.

\subsection{Scalar perturbation}
\label{subsec:scalar_pert}

Like in tensor and vector perturbation, by making $b_i$, $\beta_i$, $h_i$ vectors and $h_{ij}$ tensor zero, tedrad in (\ref{tetrad}) takes the form

\begin{align}
    e^A\text{}_\mu \rightarrow  \left(
\begin{matrix}
   1+ \alpha & 0 \\
    a \delta^I_i \partial^i \beta  & a  \delta^I_i (1+\zeta)
\end{matrix}
\right) .\label{stetrad}
\end{align}
As we will see later on, we can use the constraint equations to remove $\alpha$ and $\beta$ and thus get the quadratic action only in terms of a single variable, $\zeta$.
In the case of scalar perturbations, axial, vectorial and tensorial parts of the torsion tensor are 
\begin{align}
    T_{\rm ax}=&\,0 ,
    \\[7pt]
    T_{\rm vec}=&-\frac{9\dot{a}^2}{a^2}+\frac{6 \dot{a}}{a^2} \left[3\dot{a} \alpha +a (\bm{\nabla}^2\beta-3 \dot{\zeta}) \right] +\frac{1}{a^2} \Big\{-27 \alpha^2 \dot{a}^2+6\, a\dot{a} \left[ 3 
    \bm{\nabla}\beta \,\bm{\nabla}\zeta-2 \alpha (\bm{\nabla}^2\beta-3\,\dot{\zeta})\,\right]\,+ \\ \nonumber
    &+\left[\,(2\bm{\nabla}\zeta+\bm{\nabla}\alpha)^2-a^2(\bm{\nabla}^2\beta-3\dot{\zeta})^2\,\right]\Big\} ,
    \\[7pt]
    T_{\rm ten}=&\left[\frac{(\bm{\nabla}\zeta-\bm{\nabla}\alpha)^2}{a^2}-(\bm{\nabla}^2\beta)^2\right] .
\end{align}

After plugging perturbed tetrad Eq.~(\ref{stetrad}) into the action and expanding it to the second order, we get
\begin{equation}
   S_{\rm S}^{(2)}=\int \dd t \dd^3 x a^3 \Bigg[-3 \mathcal{A} \Dot{\zeta}^2+\frac{\mathcal{B}}{a^2}(\bm{\nabla} \zeta)^2+\Sigma \alpha^2-2 \Theta \alpha \bm{\nabla}^2 \beta+2 \mathcal{A}\dot{\zeta}\bm{\nabla}^2 \beta+6\Theta \alpha \dot{\zeta} - 2 \mathcal{C} \alpha\frac{\bm{\nabla}^2}{a^2}\zeta \Bigg] ,
   \label{s action}
\end{equation}
where the new coefficients
\begin{align}
    \mathcal{A}:=& 2\Big[G_4-2XG_{4,X}+XG_{5, \phi}-G_{\rm Tele,T}+\frac{3}{2}(G_{{\rm Tele},T_{\rm vec}}-XG_{{\rm Tele},I_2 I_2})-3H^2(4G_{{\rm Tele},TT}-12G_{{\rm Tele},TT_{\rm vec}} \nonumber   \\  
    &+9G_{{\rm Tele},T_{\rm vec} T_{\rm vec}})- H(XG_{5,X}+6G_{{\rm Tele},T I_2}-9G_{{\rm Tele},T_{\rm vec} I_2}) \dot{\phi} \Big], \\
   \mathcal{B}:=& \frac{2}{9}\Big(9 G_{4}-9 X G_{5,\phi}-6 X G_{{\rm Tele},J_{3}}-5 X G_{{\rm Tele},J_{5}}+2 X^2 G_{{\rm Tele},J_{6}} - 2 X G_{{\rm Tele},J_{8}}-9 G_{{\rm Tele},T} + 18 G_{{\rm Tele},T_{\rm vec}} \nonumber \\ 
   &-9 X G_{5,X}\ddot{\phi}\Big), \\
   \mathcal{C}:=& 2\big(G_4-2XG_{4,X}+XG_{5, \phi}-HXG_{5,X}\dot{\phi}-G_{\rm Tele,T}+G_{\rm Tele,T_{\rm vec}}\big) \nonumber\\
   &+\frac{X}{9}(3G_{\rm Tele,J_3}+10G_{\rm Tele,J_5}-4XG_{\rm Tele,J_6}+4G_{\rm Tele,J_8}) \,, \\
   \Sigma:=& X \left(G_{{\rm Tele},X}+2 X G_{{\rm Tele},XX}+2 X G_{2,XX}+G_{2,X}-2 X G_{3,\phi X}-2 G_{3,\phi}\right)+3 H \dot{\phi} \big(4 X G_{{\rm Tele},X I_{2}}+G_{{\rm Tele},I_{2}} \nonumber \\ 
   &+2 X^2 G_{3,XX}+4 X G_{3,X}-4 X^2 G_{4,\phi X X}-10 X G_{4,\phi X}-2 G_{4,\phi}\big)+3 H^2 \big(12 X G_{{\rm Tele},I_{2} I_{2}}+8 X G_{{\rm Tele},X T} \nonumber \\
   &+2 G_{{\rm Tele},T}-2 G_{4}-3 G_{{\rm Tele},T_{\rm vec}}+8 X^3 G_{4,XXX}+32 X^2 G_{4,XX}+14 X G_{4,X}-12 X G_{{\rm Tele},XT_{\rm vec}}-4 X^3 G_{5,\phi XX} \nonumber \\ 
   &-18 X^2 G_{5,\phi X}-12 X G_{5,\phi} \big)+2 H^3 \dot{\phi} \left(36 G_{{\rm Tele},TI_{2}}-54 G_{{\rm Tele},T_{\rm vec} I_{2}}+2 X^3 G_{5,XXX}+13 X^2 G_{5,XX}+15 X G_{5,X}\right) \nonumber \\ 
   &+18 H^4 \left(-12 G_{{\rm Tele},TT_{\rm vec}}+4 G_{{\rm Tele},TT}+9 G_{{\rm Tele},T_{\rm vec} T_{\rm vec}}\right), \\
    \Theta:=& -6 H^3 (4 G_{{\rm Tele},TT}-12 G_{{\rm Tele},TT_{\rm vec}}+9 G_{{\rm Tele},T_{\rm vec}T_{\rm vec}})+ H(2 G_{4}-8 X G_{4,X}-8 X^2 G_{4,XX}+6 X G_{5,\phi}\nonumber \\ 
    &+4 X^2 G_{5,\phi X}-2 G_{{\rm Tele},T}+3 G_{{\rm Tele},T_{\rm vec}}-6 X G_{{\rm Tele},I_{2}I_{2}}-4 X G_{{\rm Tele},XT}+6 X G_{{\rm Tele},XT_{\rm vec}}) \nonumber \\ 
    &- H^2 (5 X G_{5,X}+2 X^2 G_{5,XX}+18 G_{{\rm Tele},TI_{2}}-27 G_{{\rm Tele},T_{\rm vec}I_{2}}) \dot{\phi} \nonumber \\
    &-(X G_{3,X}-G_{4,\phi}-2 X G_{4,\phi X}+\frac{1}{2}G_{{\rm Tele},I_{2}}+ X G_{{\rm Tele},XI_{2}}) \dot{\phi} \, ,
\end{align}
have been introduced. The coefficients of $\zeta^2$ and $\alpha \zeta$ vanish due to the background equations. If we check this result to the limiting cases, it is easy to see that if $G_{\rm Tele} \xrightarrow[]{}0$, we can recover the same results for the action and its coefficient as in Horndeski case Ref.~\cite{Kobayashi:2011nu}. To be more precise, if we consider the vanishing of the teleparallel terms, we get $\mathcal{A}=\mathcal{C}=\mathcal{G}_{\rm T}$ and $\mathcal{B}=\mathcal{F}_{\rm T}$.

By varying the action \eqref{s action} with respect to $\alpha$ and $\beta$, one obtains a set of equations
\begin{gather}
    \Sigma \alpha-\Theta \bm{\nabla}^2\beta+3 \Theta \dot{\zeta}-\mathcal{C}\frac{\bm{\nabla}^2}{a^2}\zeta = 0 \,,   \\
    \Theta\alpha-\mathcal{A}\dot{\zeta} = 0\, .
\end{gather}
Using these equations, we can eliminate $\alpha$ and $\beta$ from the action \eqref{s action} and rewrite it as
\begin{align}\label{quadratic scalar action}
     S_{\rm S}^{(2)}=\int \dd t \dd^3 x a^3 \Bigg[\mathcal{G_S}\dot{\zeta}^2-\frac{\mathcal{F_S}}{a^2}(\bm{\nabla}\zeta)^2\Bigg] ,
\end{align}
where the new coefficients are 
\begin{gather}
\mathcal{G}_{\rm S} = 3  \mathcal{A}+\frac{\Sigma \mathcal{A}^2}{\Theta^2} \,,     \\
    \mathcal{F}_{\rm S} = \frac{1}{a}\frac{d}{dt}\Bigg(\frac{a  \mathcal{A}\mathcal{C}}{\Theta}\Bigg)-\mathcal{B} \,.
\end{gather}

We can also express $\mathcal{A}$ , $\mathcal{B}$ and $\mathcal{C}$ coefficients with $\mathcal{F}_{\rm T}$, $\mathcal{G}_{\rm T}$ and $G_{\rm Tele}$ like
\begin{align}
    \mathcal{A}&=\mathcal{G}_{\rm T}+f_1\left(G_{\rm Tele}\right),
    \\
     \mathcal{B}&=\mathcal{F}_{\rm T}+ f_2\left(G_{\rm Tele}\right),
    \\
    \mathcal{C}&=\mathcal{G}_{\rm T}+f_3\left(G_{\rm Tele}\right).
\end{align}
where
\begin{align}
    f_1\left(G_{\rm Tele}\right)&=3G_{\rm Tele,T_{\rm vec}}-X\left(G_{\rm Tele,J_5}+G_{\rm Tele,J_8}+3G_{\rm Tele,I_2 I_2}\right)+6H \dot{\phi}\left(3G_{\rm Tele,T_{\rm vec}I_2}-2G_{\rm Tele,TI_2}\right)+ \\ \nonumber
    &+6H^2\left(12G_{\rm Tele,TT_{\rm vec}}-4G_{\rm Tele,TT}-9G_{\rm Tele,T_{\rm vec}T_{\rm vec}}\right),
    \\ 
    f_2\left(G_{\rm Tele}\right)&=\frac{2}{9}\left(18G_{\rm Tele,T_{\rm vec}}-6XG_{\rm Tele,J_3}-5XG_{\rm Tele,J_5}-2XG_{\rm Tele,J_8}+2X^2G_{\rm Tele,J_6}\right),
    \\ 
     f_3\left(G_{\rm Tele}\right)&=\frac{1}{9}\left(18G_{\rm Tele,T_{\rm vec}}+3XG_{\rm Tele,J_3}+XG_{\rm Tele,J_5}-32XG_{\rm Tele,J_8}-4X^2G_{\rm Tele,J_6}  \right).
\end{align}
We can write $\mathcal{G}_S$ and $\mathcal{F}_S$ in terms of $\mathcal{G}_T$ and $\mathcal{F}_T$ as follows
\begin{align}
   \mathcal{G}_{\rm S} &= 3 \mathcal{G}_T+\frac{\Sigma}{\Theta^2}\left[\mathcal{G}_{\rm T}+f_1\left(G_{\rm Tele}\right)\right]^2 +3f_1(G_{\rm Tele})\,,     \\
   \mathcal{F}_{\rm S} &= \frac{1}{a}\frac{d}{dt}\Bigg \{\frac{a \left[\mathcal{G}_{\rm T}+f_1\left(G_{\rm Tele}\right)\right] \left[\mathcal{G}_{\rm T}+f_3\left(G_{\rm Tele}\right)\right]}{\Theta}\Bigg\}-\mathcal{F}_T-f_2(G_{\rm Tele}) \,.
\end{align}

The square sound speed is given by $c_{\rm S}^2=\mathcal{F}_{\rm S}/\mathcal{G}_{\rm S}$ and ghost and gradient stability are obtained
\begin{align}
    \mathcal{F}_{\rm S}>0, \quad {\rm }\quad \mathcal{G}_{\rm S}>0 .
\end{align}

These considerations are essential for understanding the speed at which scalar perturbations propagate from the early Universe which can impact the predicted size of baryonic acoustic oscillations.

\section{Applications}\label{sec:application}

The results we found in the above analysis can be used in several applications in cosmology, especially in the early Universe. In order to understand the formation and evolution of large-scale structure one has to understand the nature of cosmological perturbations. In this section, we will discuss the power spectrum of the perturbations above, and we will provide the form of the alpha parameters that can be used to distinguish between the concordance model and other alternatives. 

\subsection{Primordial power spectrum}

The seeds of all structure in the Universe are considered to be the primordial fluctuations, whose origin is most probably related to inflation. Before the freezing of the horizon at the very early Universe, as the scale factor grew exponentially, it caused quantum fluctuations of the inflaton field, which at later stages re-entered the horizon and set the initial conditions for the formation of large scale structure. These fluctuations are usually described by their power spectrum that has both scalar and tensor modes. That is what we discuss in this section. 

\paragraph{Tensor perturbations}

The quadratic action \eqref{eq:ten_pert_action} in tensor perturbations can be re-written using the canonical variables
\begin{equation}
    \dd y _{\rm T} := \frac{\mathcal{F}^{1/2}_{\rm T}}{a\mathcal{G}_{\rm T}} \dd t\,,\,\,\, z_{\rm T} := \frac{a}{2} (\mathcal{F}_{\rm T} \mathcal{G}_{\rm T})^{1/4}\,,\,\,\, v _{ij}(y_{\rm T},\bm{x}) := z_{\rm T} h_{ij}\,, 
\end{equation}
as
\begin{equation}
    \mathcal{S}_{\rm T} ^{(2)} = \int \dd y_{\rm T}\dd ^3x \left[ (v' _{ij})^2 - (\bm{\nabla} v_{ij})^2 + \frac{z_{\rm T}''}{z_{\rm T}} v_{ij}^2 \right]\,,
\end{equation}
where prime denotes differentiation with respect to $y_{\rm T}$. Varying this action with respect to $v_{ij}$ and solving its equation, we see that at superhorizon scales we get
\begin{equation}
    v_{ij} \propto z_{\rm T} \,,\quad  v_{ij} \propto z_{\rm T} \int \frac{\dd y_{\rm T}}{z^2_{\rm T}}\,, \label{canonical sol}
\end{equation}
or in terms of the non-canonical variables
\begin{equation}
    h_{ij} = {\rm const} \,, \quad h_{ij} = \int ^t \frac{\dd t'}{a^3 \mathcal{G}_{\rm T}}\,. \label{non canonical sol}
\end{equation}

To evaluate the power spectral density, we assume that
\begin{equation}\label{assumptions}
    \epsilon := - \frac{\dot{H}}{H^2} \simeq {\rm const}\,,\,\, f_{\rm T}:= \frac{\dot{\mathcal{F}}_{\rm T}}{H \mathcal{F}_{\rm T}} \simeq {\rm const}\,,\,\, g_{\rm T} := \frac{\dot{\mathcal{G}}_{\rm T}}{H \mathcal{G}_{\rm T}} \simeq {\rm const}\,.
\end{equation}
In order for the canonical time coordinate to run from $- \infty$ to $0$ as the Universe expands, we have to impose the condition
\begin{equation}\label{condition-1}
    \epsilon + \frac{f_{\rm T} - g_{\rm T}}{2} < 1\,,
\end{equation}
while in order for the second solution in Eq.~\eqref{canonical sol}-\eqref{non canonical sol} to decay, we have to assume that
\begin{equation}\label{condition-2}
    \epsilon - g_{\rm T} < 3\,.
\end{equation}

Expanding the tensor perturbations in terms of the eigenfunctions $e^{i \bm{k}\cdot \bm{x}}$ of the Laplacian and the polarization tensor $\mathrm{e}_{ij}$ in the Fourier space, we can solve the mode functions equation to get
\begin{equation}
    v_{ij} = \frac{\sqrt{\pi}}{2} \sqrt{-y_{\rm T}}H_{\nu _{\rm T}}^{(1)} (-k y_{\rm T})\mathrm{e}_{ij}\,,
\end{equation}
with $H_{\nu _{\rm T}}^{(1)}$ being the Hankel function of first kind (plus sign) and  $\nu _{\rm T}$ being a positive scalar defined as
\begin{equation}
    \nu _{\rm T} := \frac{3 - \epsilon + g_{\rm T}}{2 - 2 \epsilon - f_{\rm T} + g_{\rm T}}\,.
\end{equation}
Thus, the power spectrum of the primordial tensor fluctuations becomes
\begin{equation}
    \mathcal{P}_{\rm T} = 8 \gamma _{\rm T}  \frac{\mathcal{G}_{\rm T}^{1/2}}{\mathcal{F}_{\rm T} ^{3/2}} \frac{H^2}{ 4\pi ^2} \Big\rvert _{-k y_{\rm T} = 1}\,,
\end{equation}
where $$
\gamma _{\rm T} = 2^{2\nu_{\rm T} - 3}  \left|\frac{\Gamma (\nu _{T})}{\Gamma (3/2)}\right|^2 (1-\epsilon - \frac{f_{\rm T}}{2} + \frac{g_{\rm T}}{2})\,.$$
We evaluate the power spectrum at the sound horizon exit, i.e. $-ky_{\rm T} = 1$, since $c_{\rm T} \neq c$ in general. The tensor spectral index is given by
\begin{equation}\label{tensor spectral index}
    n_{\rm T} = 3 - 2 \nu _{\rm T}\,,
\end{equation}
and the scale-invariant limit for tensor perturbations would be for $\nu _{\rm T} = 3/2$. From Eq.~\eqref{tensor spectral index} we see that, the gravitational wave spectrum could have a blue tilt if
\begin{equation}
    n_{\rm T} > 0 \Rightarrow 4 \epsilon + 3 f_{\rm T} - g_{\rm T} < 0\,.
\end{equation} 
Notice that the conditions \eqref{condition-1} and \eqref{condition-2} are not affected by this and even if B-mode polarization were to be detected on CMB, the theory could still be a good candidate. However, we should mention that the assumptions \eqref{assumptions} may not always hold, as long as the above conditions are satisfied. 

\paragraph{Scalar perturbations}

In order to canonically normalize the action \eqref{quadratic scalar action} we define the following variables
\begin{equation}
    \dd y_{\rm S} := \frac{\mathcal{F}_{\rm S}^{1/2}}{a\mathcal{G}_{\rm S}^{1/2}} \dd t \,,\,\,\, z_{\rm S} := \sqrt{2}a (\mathcal{F}_{\rm S}\mathcal{G}_{\rm S})^{1/4} \,,\,\,\, u(y_{\rm S},\bm{x}) := z_{\rm S} \zeta\,.
\end{equation}
Plugging them in the quadratic action we get
\begin{equation}
    \mathcal{S}_{\rm S}^{(2)} =  \frac{1}{2} \int \dd y_{\rm S} \dd ^3 x \left[ (u')^2 - (\bm{\nabla}u)^2 + \frac{z'' _{\rm S}}{z_{\rm S}}u^2 \right]\,,
\end{equation}
where, as previously, prime denotes differentiation with respect to the canonical time variable, $y_{\rm S}$.

Following the same procedure as with the tensor perturbations to evaluate the power spectrum in this case, we assume
\begin{equation}
   \epsilon := - \frac{\dot{H}}{H^2} \simeq {\rm const}\,,\,\, f_{\rm S}:= \frac{\dot{\mathcal{F}}_{\rm S}}{H \mathcal{F}_{\rm S}} \simeq {\rm const}\,,\,\, g_{\rm S} := \frac{\dot{\mathcal{G}}_{\rm S}}{H \mathcal{G}_{\rm S}} \simeq {\rm const}\,.
\end{equation}
Then, the power spectrum will be given by
\begin{equation}
    \mathcal{P}_{\rm S} = \frac{\gamma _{\rm S}}{2} \frac{\mathcal{G}_{\rm S}^{1/2}}{\mathcal{F}_{\rm S}^{3/2}} \frac{H^2}{4\pi ^2}\Big\rvert_{-ky_{\rm S} = 1}\,,
\end{equation}
where $$\nu _{\rm S}:= \frac{3 - \epsilon + g_{\rm S}}{2- 2\epsilon - f_{\rm S} + g_{\rm S}}$$ and $$\gamma _{\rm S} = 2 ^{2 \nu _{\rm S}-3} \left|\frac{\Gamma (\nu _{\rm S})}{\Gamma (3/2)}\right|^2(1-\epsilon - \frac{f_{\rm S}}{2}+\frac{g_{\rm S}}{2})\,.$$ 
The scalar spectral index is given by
\begin{equation}
    n_{\rm S} = 4 - 2 \nu _{\rm S}\,,
\end{equation}
and thus a spectrum with equal amplitudes at horizon crossing should obey
\begin{equation}
    \epsilon + \frac{3 f_{\rm S}}{4} - \frac{g_{\rm S}}{4} = 0\,.
\end{equation}

If we consider the limit $\epsilon, f_{\rm T}, g_{\rm T}, f_{\rm S},g_{\rm S} \ll 1$, we get $\nu _{\rm T} , \nu _{\rm S} \rightarrow 3/2$ and thus $\gamma _{\rm T} , \gamma _{\rm S} \rightarrow 1$. The tensor-to-scalar ration is then given by
\begin{equation}
    r = 16 \frac{\mathcal{F}_{\rm S}}{\mathcal{F}_{\rm T}} \frac{c_{\rm S}}{c _{\rm T}}\,.
\end{equation}
Given that, one can consider any  model of inflation that is a sub-class of the BDLS theory and find the form $r$ in terms of the slow-roll parameters.

\subsection{Alpha parameters}

It has been proposed in \cite{Bellini:2014fua} that perturbation dynamics can be solely described by four functions, $\alpha _i$. Similarly to effective field theory methods, one can use observations to constrain the value of these four parameters, without specifying any particular model or initial conditions and thus test possible deviations from $\Lambda$CDM. 

Given a specific background evolution, which can also be obtained merely by the observations \cite{Dialektopoulos:2023dhb,Dialektopoulos:2023jam,Briffa:2023ern,Dialektopoulos:2021wde,LeviSaid:2021yat,Hwang:2022hla}, and the value of the matter density today \cite{Planck:2018vyg}, one can fully determine the evolution of large-scale structure in the Universe. The physical meaning of these time-dependent functions is as follows:
\begin{itemize}
    \item \textit{Kineticity:} $\alpha _{\rm K}$. Indicates the kinetic energy of curvature perturbations. It is essentially unconstrained by observations since it has no actual impact on any of the observables and large values tend to suppress the sound speed of the perturbations. The contributions for $\alpha _{\rm K}$ are from all the $G_2, G_3, G_4, G_5$ and $G_{\rm Tele}$.
    \item \textit{Braiding:} $\alpha _{\rm B}$. Its presence shows evidence of the mixing between the kinetic term of the scalar field and the metric. As $\alpha _{\rm K}$, it only affects the curvature perturbations, giving rise to a quintessence-like force. It is the reason for dark energy clustering and has contributions from $G_{\rm Tele}$ and all $G_i$ functions except the potential-like term $G_2$. 
    \item \textit{Planck mass run rate:} $\alpha _{\rm M}$. Merely a redefinition of the Planck mass that does not affect physics. It contributes to both curvature and tensor perturbations and also creates anisotropic stress in the Jordan frame. It receives contributions from $G_4, G_5$ and $G_{\rm Tele}$ terms.
    \item \textit{Tensor speed excess:} $\alpha _{\rm T}$. Describes the propagation speed of gravitational waves through the tensor perturbations and specifically its deviation from the speed of light. This is done through the relation $c_{\rm T}^2 = 1+ \alpha _{\rm T}$ their values could be constrained by cosmological observations and experiments. Its contributions are from $G_4, G_5$ and $G_{\rm Tele}$.
\end{itemize}

As already seen in \cite{Saltas:2014dha,Riazuelo:2000fc} the most general parametrization of the gravitational wave perturbation equation on a flat cosmological background in modified gravity takes the form
\begin{equation}
    \ddot{h}_{ij} + (3 + \alpha _{\rm M}) H \dot{h}_{ij} - (1+ \alpha _{\rm T}) \frac{k}{a^2}h_{ij} = 0\,,
\end{equation}
where dots are derivatives with respect to cosmic time and $k$ the wavenumber of the perturbation on the Fourier space. 

In BDLS theory the excess tensor speed parameter is given by \cite{Bahamonde:2019ipm}
\begin{equation}
     M_{\ast}^2 a_{\rm T} \equiv 2 X \left(  2G_{4,X}-2 G_{5,\phi}-(\ddot{\phi}-H\dot{\phi})G_{5,X}-2G_{{\rm Tele}, J_8}-\frac{1}{2}G_{{\rm Tele}, J_5}\right)\,,
\end{equation}
and the effective Planck mass by
\begin{equation}
  M_{\ast}^2 \equiv 2\left(G_4-2 X G_{4,X}+  X G_{5,\phi}-H X \dot{\phi} G_{5,X} + 2 X G_{{\rm Tele}, J_8} + \frac{1}{2}XG_{{\rm Tele}, J_5} - G_{{\rm Tele},T}\right)\,.
\end{equation}
In addition, the Planck mass run rate parameter can be determined by
\begin{equation}
    H M_{\ast}^2 \alpha _{\rm M} \equiv \frac{\dd M_{\ast}^2}{\dd t}\,. 
\end{equation}
The remaining two alpha-parameters, i.e. braiding and kineticity, are expressed respectively as
\begin{align}
    \mathcal{A}H \alpha_{B}=& 2\dot{\phi}(X G_{3X}-G_{4\phi}-2X G_{4\phi X})+2XH(4G_{4X}+8XG_{4XX}-4G_{5\phi}-4XG_{5\phi X}+\\ \nonumber
    &+3G_{{\rm Tele},I_2I_2}+4G_{{\rm Tele},XT}-6G_{{\rm Tele},XT_{\rm vec}})+2 \dot{\phi} X H^2(3 G_{5X}+2XG_{5XX})+ \\ \nonumber
    &+\dot{\phi}(G_{\rm Tele,I_2}+12H^2 G_{{\rm Tele},TI_2}-18H^2 G_{{\rm Tele},T_{\rm vec} I_2}+2X G_{{\rm Tele},X I_2}),
\end{align}
\begin{align}
    \mathcal{A}H^2 \alpha_{K}=& 2X(G_{2X}+2XG_{2XX}-2G_{3\phi}-2XG_{3\phi X})+12\dot{\phi} X H(G_{3X}+XG_{3XX}-3G_{4 \phi X}-2XG_{4\phi X X})+\\ \nonumber
    &+12XH^2(G_{4X}+8XG_{4XX}+4X^2G_{4XXX}-G_{5\phi}-5XG_{5\phi X}-2X^2G_{5\phi XX})+ \\ \nonumber
    &+4\dot{\phi}XH^3(3G_{5X}+7XG_{5XX}+2X^2G_{5XXX})+2X(9H^2G_{{\rm Tele},I_2I_2}+2XG_{{\rm Tele},XX}+6\dot{\phi}HG_{{\rm Tele},XI_2}+G_{{\rm Tele},X}).
\end{align}
Notice that in the $G_{\rm Tele}\rightarrow 0$ limit, $\mathcal{A} \rightarrow M_{\ast}$ and one can derive the respective expressions in the standard Horndeski cosmology~\cite{Bellini}. The presence of the teleparallel contribution here, in terms of $G_{\rm Tele}$ might seemingly complicate things, but it also adds interesting phenomenology in the structure of the theory. This can become evident from the fact that functions that were severely constrained in the standard Horndeski formulation, such as the $G_{4,X}$ and $G_5$, can survive in this setup.

Furthermore, one can rewrite the expressions in the action \eqref{quadratic scalar action} in terms of the alpha parameters and obtain
\begin{gather}
     \Theta=\frac{\mathcal{A} H}{2} (2-\alpha_{B}) \,\, {\rm and}\,\, \Sigma=-\frac{\mathcal{A}H^2}{2}( 6-\alpha_{K}-6 \alpha_{B}), 
    \\ \nonumber
    \mathcal{G_S}=\frac{2 \mathcal{A} D}{(2-\alpha_B)^2}, \,\, \rm where \,\,D=\alpha_K+\frac{3}{2} \alpha_B^2.
\end{gather}
If we consider $\mathcal{C}H\alpha_X=d\mathcal{C}/dt$ equation holds for $\alpha_X$ parameter, squared sound speed can be found with the following formula
\begin{align}
    c_s^2&=\frac{\mathcal{C}}{\mathcal{A}}\,\,\frac{(2-\alpha_B)(H^2(1+\alpha_X)-\dot{H})+\dot{\alpha_B}H}{D H^2}-\frac{\mathcal{B}}{\mathcal{A}}\,\, \frac{(2-\alpha_B)^2}{2D}.
\end{align}

Here we introduced a new parameter $\alpha_{X}$ by using the similar approach like $\alpha_{M}$ in Horndeski gravity, in order to give more compact form to the formula of squared sound speed. Also, as it is mentioned earlier, if we make teleparallel terms zero ($G_{\rm Tele} \xrightarrow[]{}0$), then $\mathcal{A}=\mathcal{C}=\mathcal{G}_T$ and $\mathcal{B}=\mathcal{F}_T$ which means that we get the same expression for squared sound speed like in Horndeski case.


\section{Conclusions} \label{sec:conc}

A homogeneous and isotropic FLRW Universe is the pedestal of modern cosmology, even if it has been argued otherwise~\cite{Krishnan:2021jmh}. It is on top of this background that the cosmological perturbations propagate and become source for many of our observations. 

In this paper, we study in detail the cosmological perturbations around a flat FLRW spacetime in the teleparallel analog of Horndeski gravity. It has been shown before \cite{Bahamonde:2019shr}, that BDLS theory presents much richer phenomenology compared to the standard/curvature Horndeski, because of the presence of $G_{\rm Tele}$ function, which depends on all those teleparallel quantities that in pure Riemannian geometry do not exist. Specifically, it was found that \cite{Bahamonde:2019ipm} in BDLS theory, terms like $G_{4,X}$ and $G_5(\phi,X)$ can survive, in contrast with the standard Horndeski formulation in Riemannian geometry, where they are severely constrained (if not eliminated) because they predict a different speed for the propagation of gravitational waves. 

Here we present the scalar, vector and tensor perturbations in this theory and we show how the propagation speed of both tensor and scalar perturbations is affected by the presence of the extra teleparallel Lagrangian. In addition, after normalizing the quadratic scalar and tensor perturbations action, we switch to the Fourier space and calculate the power spectra of primordial fluctuations. We also express the tensor-to-scalar ratio in terms of the perturbations coefficients in the action. In this way, one could assume a specific teleparallel Horndeski model, fix the $G_i$ and $G_{\rm Tele}$ functions and calculate immediately the tensor-to-scalar ratio. 

Last but not least, we present the perturbations in terms of the so-called \textit{alpha} parameters, that is the four parameters that one can constrain solely from observations without the need to specify any physical model. In this way, we could possibly discriminate between the concordance cosmological model and alternative descriptions.

As far as future projects are concerned, it would be very interesting to see how this analysis is affected in a cosmological spacetime with non-trivial spatial curvature. Furthermore, we plan to investigate possible evasions of the no-go theorem that manifests itself in the Riemannian Horndeski gravity regarding bouncing solutions.


\begin{acknowledgments}
The work was supported by the PNRR-III-C9-2022–I9 call, with project number 760016/27.01.2023, by the Nazarbayev University Faculty Development Competitive Research Grant No. 11022021FD2926 and by the Hellenic Foundation for Research and Innovation (H.F.R.I.) under the “First Call for H.F.R.I. Research Projects to support Faculty members and Researchers and the procurement of high-cost research equipment grant” (Project Number: 2251). This article is also based upon work from COST Action CA21136 Addressing observational tensions in cosmology with systematics and fundamental physics (CosmoVerse) supported by COST (European Cooperation in Science and Technology).
\end{acknowledgments}

\bibliographystyle{utphys}
\bibliography{references}

\providecommand{\href}[2]{#2}\begingroup\raggedright\begin{thebibliography}{10}

\bibitem{SupernovaSearchTeam:1998fmf}
{\bf Supernova Search Team} Collaboration, A.~G. Riess {\em et al.},
  ``{Observational evidence from supernovae for an accelerating universe and a
  cosmological constant},'' \href{http://dx.doi.org/10.1086/300499}{{\em
  Astron. J.} {\bf 116} (1998)  1009--1038},
  \href{http://arxiv.org/abs/astro-ph/9805201}{{\tt arXiv:astro-ph/9805201}}.

\bibitem{SupernovaCosmologyProject:1998vns}
{\bf Supernova Cosmology Project} Collaboration, S.~Perlmutter {\em et al.},
  ``{Measurements of $\Omega$ and $\Lambda$ from 42 high redshift
  supernovae},'' \href{http://dx.doi.org/10.1086/307221}{{\em Astrophys. J.}
  {\bf 517} (1999)  565--586},
  \href{http://arxiv.org/abs/astro-ph/9812133}{{\tt arXiv:astro-ph/9812133}}.

\bibitem{DiValentino:2021izs}
E.~Di~Valentino, O.~Mena, S.~Pan, L.~Visinelli, W.~Yang, A.~Melchiorri, D.~F.
  Mota, A.~G. Riess, and J.~Silk, ``{In the Realm of the Hubble tension $-$ a
  Review of Solutions},'' \href{http://arxiv.org/abs/2103.01183}{{\tt
  arXiv:2103.01183 [astro-ph.CO]}}.

\bibitem{Peebles:2002gy}
P.~J.~E. Peebles and B.~Ratra, ``{The Cosmological Constant and Dark Energy},''
  \href{http://dx.doi.org/10.1103/RevModPhys.75.559}{{\em Rev. Mod. Phys.} {\bf
  75} (2003)  559--606}, \href{http://arxiv.org/abs/astro-ph/0207347}{{\tt
  arXiv:astro-ph/0207347}}.

\bibitem{Copeland:2006wr}
E.~J. Copeland, M.~Sami, and S.~Tsujikawa, ``{Dynamics of dark energy},''
  \href{http://dx.doi.org/10.1142/S021827180600942X}{{\em Int. J. Mod. Phys. D}
  {\bf 15} (2006)  1753--1936}, \href{http://arxiv.org/abs/hep-th/0603057}{{\tt
  arXiv:hep-th/0603057}}.

\bibitem{Weinberg:1988cp}
S.~Weinberg, ``{The Cosmological Constant Problem},''
  \href{http://dx.doi.org/10.1103/RevModPhys.61.1}{{\em Rev. Mod. Phys.} {\bf
  61} (1989)  1--23}.

\bibitem{Gaitskell:2004gd}
R.~J. Gaitskell, ``{Direct detection of dark matter},''
  \href{http://dx.doi.org/10.1146/annurev.nucl.54.070103.181244}{{\em Ann. Rev.
  Nucl. Part. Sci.} {\bf 54} (2004)  315--359}.

\bibitem{DiValentino:2020vhf}
E.~Di~Valentino {\em et al.}, ``{Snowmass2021 - Letter of interest cosmology
  intertwined I: Perspectives for the next decade},''
  \href{http://dx.doi.org/10.1016/j.astropartphys.2021.102606}{{\em Astropart.
  Phys.} {\bf 131} (2021)  102606}, \href{http://arxiv.org/abs/2008.11283}{{\tt
  arXiv:2008.11283 [astro-ph.CO]}}.

\bibitem{DiValentino:2020zio}
E.~Di~Valentino {\em et al.}, ``{Cosmology Intertwined II: The Hubble Constant
  Tension},'' \href{http://arxiv.org/abs/2008.11284}{{\tt arXiv:2008.11284
  [astro-ph.CO]}}.

\bibitem{DiValentino:2020vvd}
E.~Di~Valentino {\em et al.}, ``{Cosmology Intertwined III: $f \sigma_8$ and
  $S_8$},'' \href{http://dx.doi.org/10.1016/j.astropartphys.2021.102604}{{\em
  Astropart. Phys.} {\bf 131} (2021)  102604},
  \href{http://arxiv.org/abs/2008.11285}{{\tt arXiv:2008.11285 [astro-ph.CO]}}.

\bibitem{Riess:2021jrx}
A.~G. Riess {\em et al.}, ``{A Comprehensive Measurement of the Local Value of
  the Hubble Constant with 1 km/s/Mpc Uncertainty from the Hubble Space
  Telescope and the SH0ES Team},'' \href{http://arxiv.org/abs/2112.04510}{{\tt
  arXiv:2112.04510 [astro-ph.CO]}}.

\bibitem{Wong:2019kwg}
K.~C. Wong {\em et al.}, ``{H0LiCOW \textendash{} XIII. A 2.4 per cent
  measurement of H0 from lensed quasars: 5.3\ensuremath{\sigma} tension between
  early- and late-Universe probes},''
  \href{http://dx.doi.org/10.1093/mnras/stz3094}{{\em Mon. Not. Roy. Astron.
  Soc.} {\bf 498} (2020) no.~1, 1420--1439},
  \href{http://arxiv.org/abs/1907.04869}{{\tt arXiv:1907.04869 [astro-ph.CO]}}.

\bibitem{Anderson:2023aga}
R.~I. Anderson, N.~W. Koblischke, and L.~Eyer, ``{Reconciling astronomical
  distance scales with variable red giant stars},''
  \href{http://arxiv.org/abs/2303.04790}{{\tt arXiv:2303.04790 [astro-ph.CO]}}.

\bibitem{Freedman:2020dne}
W.~L. Freedman, B.~F. Madore, T.~Hoyt, I.~S. Jang, R.~Beaton, M.~G. Lee,
  A.~Monson, J.~Neeley, and J.~Rich, ``{Calibration of the Tip of the Red Giant
  Branch (TRGB)},'' \href{http://arxiv.org/abs/2002.01550}{{\tt
  arXiv:2002.01550 [astro-ph.GA]}}.

\bibitem{Aghanim:2018eyx}
{\bf Planck} Collaboration, N.~Aghanim {\em et al.}, ``{Planck 2018 results.
  VI. Cosmological parameters},''
  \href{http://dx.doi.org/10.1051/0004-6361/201833910}{{\em Astron. Astrophys.}
  {\bf 641} (2020)  A6}, \href{http://arxiv.org/abs/1807.06209}{{\tt
  arXiv:1807.06209 [astro-ph.CO]}}. [Erratum: Astron.Astrophys. 652, C4
  (2021)].

\bibitem{DES:2021wwk}
{\bf DES} Collaboration, T.~M.~C. Abbott {\em et al.}, ``{Dark Energy Survey
  Year 3 results: Cosmological constraints from galaxy clustering and weak
  lensing},'' \href{http://dx.doi.org/10.1103/PhysRevD.105.023520}{{\em Phys.
  Rev. D} {\bf 105} (2022) no.~2, 023520},
  \href{http://arxiv.org/abs/2105.13549}{{\tt arXiv:2105.13549 [astro-ph.CO]}}.

\bibitem{eBOSS:2020yzd}
{\bf eBOSS} Collaboration, S.~Alam {\em et al.}, ``{Completed SDSS-IV extended
  Baryon Oscillation Spectroscopic Survey: Cosmological implications from two
  decades of spectroscopic surveys at the Apache Point Observatory},''
  \href{http://dx.doi.org/10.1103/PhysRevD.103.083533}{{\em Phys. Rev. D} {\bf
  103} (2021) no.~8, 083533}, \href{http://arxiv.org/abs/2007.08991}{{\tt
  arXiv:2007.08991 [astro-ph.CO]}}.

\bibitem{Zhang:2021yna}
P.~Zhang, G.~D'Amico, L.~Senatore, C.~Zhao, and Y.~Cai, ``{BOSS Correlation
  Function analysis from the Effective Field Theory of Large-Scale
  Structure},'' \href{http://dx.doi.org/10.1088/1475-7516/2022/02/036}{{\em
  JCAP} {\bf 02} (2022) no.~02, 036},
  \href{http://arxiv.org/abs/2110.07539}{{\tt arXiv:2110.07539 [astro-ph.CO]}}.

\bibitem{Cooke:2017cwo}
R.~J. Cooke, M.~Pettini, and C.~C. Steidel, ``{One Percent Determination of the
  Primordial Deuterium Abundance},''
  \href{http://dx.doi.org/10.3847/1538-4357/aaab53}{{\em Astrophys. J.} {\bf
  855} (2018) no.~2, 102}, \href{http://arxiv.org/abs/1710.11129}{{\tt
  arXiv:1710.11129 [astro-ph.CO]}}.

\bibitem{Krishnan:2021jmh}
C.~Krishnan, R.~Mohayaee, E.~O. Colg\'ain, M.~M. Sheikh-Jabbari, and L.~Yin,
  ``{Hints of FLRW breakdown from supernovae},''
  \href{http://dx.doi.org/10.1103/PhysRevD.105.063514}{{\em Phys. Rev. D} {\bf
  105} (2022) no.~6, 063514}, \href{http://arxiv.org/abs/2106.02532}{{\tt
  arXiv:2106.02532 [astro-ph.CO]}}.

\bibitem{Krishnan:2021dyb}
C.~Krishnan, R.~Mohayaee, E.~O. Colg\'ain, M.~M. Sheikh-Jabbari, and L.~Yin,
  ``{Does Hubble tension signal a breakdown in FLRW cosmology?},''
  \href{http://dx.doi.org/10.1088/1361-6382/ac1a81}{{\em Class. Quant. Grav.}
  {\bf 38} (2021) no.~18, 184001}, \href{http://arxiv.org/abs/2105.09790}{{\tt
  arXiv:2105.09790 [astro-ph.CO]}}.

\bibitem{Poulin:2023lkg}
V.~Poulin, T.~L. Smith, and T.~Karwal, ``{The Ups and Downs of Early Dark
  Energy solutions to the Hubble tension: a review of models, hints and
  constraints circa 2023},'' \href{http://arxiv.org/abs/2302.09032}{{\tt
  arXiv:2302.09032 [astro-ph.CO]}}.

\bibitem{DiValentino:2021imh}
E.~Di~Valentino and A.~Melchiorri, ``{Neutrino Mass Bounds in the Era of
  Tension Cosmology},'' \href{http://dx.doi.org/10.3847/2041-8213/ac6ef5}{{\em
  Astrophys. J. Lett.} {\bf 931} (2022) no.~2, L18},
  \href{http://arxiv.org/abs/2112.02993}{{\tt arXiv:2112.02993 [astro-ph.CO]}}.

\bibitem{DiValentino:2021rjj}
E.~Di~Valentino, S.~Gariazzo, C.~Giunti, O.~Mena, S.~Pan, and W.~Yang,
  ``{Minimal dark energy: Key to sterile neutrino and Hubble constant
  tensions?},'' \href{http://dx.doi.org/10.1103/PhysRevD.105.103511}{{\em Phys.
  Rev. D} {\bf 105} (2022) no.~10, 103511},
  \href{http://arxiv.org/abs/2110.03990}{{\tt arXiv:2110.03990 [astro-ph.CO]}}.

\bibitem{CANTATA:2021ktz}
{\bf CANTATA} Collaboration, E.~N. Saridakis {\em et al.}, ``{Modified Gravity
  and Cosmology: An Update by the CANTATA Network},''
  \href{http://arxiv.org/abs/2105.12582}{{\tt arXiv:2105.12582 [gr-qc]}}.

\bibitem{Addazi:2021xuf}
A.~Addazi {\em et al.}, ``{Quantum gravity phenomenology at the dawn of the
  multi-messenger era\textemdash{}A review},''
  \href{http://dx.doi.org/10.1016/j.ppnp.2022.103948}{{\em Prog. Part. Nucl.
  Phys.} {\bf 125} (2022)  103948}, \href{http://arxiv.org/abs/2111.05659}{{\tt
  arXiv:2111.05659 [hep-ph]}}.

\bibitem{BeltranJimenez:2019esp}
J.~Beltr\'an~Jim\'enez, L.~Heisenberg, and T.~S. Koivisto, ``{The Geometrical
  Trinity of Gravity},'' \href{http://dx.doi.org/10.3390/universe5070173}{{\em
  Universe} {\bf 5} (2019) no.~7, 173},
  \href{http://arxiv.org/abs/1903.06830}{{\tt arXiv:1903.06830 [hep-th]}}.

\bibitem{Hehl:1994ue}
F.~W. Hehl, J.~D. McCrea, E.~W. Mielke, and Y.~Ne{'{e}}man, ``{Metric affine
  gauge theory of gravity: Field equations, Noether identities, world spinors,
  and breaking of dilation invariance},''
  \href{http://dx.doi.org/10.1016/0370-1573(94)00111-F}{{\em Phys. Rept.} {\bf
  258} (1995)  1--171},
\href{http://arxiv.org/abs/gr-qc/9402012}{{\tt arXiv:gr-qc/9402012 [gr-qc]}}.

\bibitem{Aldrovandi:2013wha}
R.~Aldrovandi and J.~G. Pereira,
  \href{http://dx.doi.org/10.1007/978-94-007-5143-9}{{\em {Teleparallel
  Gravity}}}, vol.~173.
\newblock Springer, Dordrecht,
2013.
\newblock

\bibitem{Bahamonde:2021gfp}
S.~Bahamonde, K.~F. Dialektopoulos, C.~Escamilla-Rivera, G.~Farrugia, V.~Gakis,
  M.~Hendry, M.~Hohmann, J.~L. Said, J.~Mifsud, and E.~Di~Valentino,
  ``{Teleparallel Gravity: From Theory to Cosmology},''
  \href{http://arxiv.org/abs/2106.13793}{{\tt arXiv:2106.13793 [gr-qc]}}.

\bibitem{Krssak:2018ywd}
M.~Krssak, R.~van~den Hoogen, J.~Pereira, C.~Böhmer, and A.~Coley,
  ``{Teleparallel theories of gravity: illuminating a fully invariant
  approach},'' \href{http://dx.doi.org/10.1088/1361-6382/ab2e1f}{{\em Class.
  Quant. Grav.} {\bf 36} (2019) no.~18, 183001},
  \href{http://arxiv.org/abs/1810.12932}{{\tt arXiv:1810.12932 [gr-qc]}}.

\bibitem{Cai:2015emx}
Y.-F. Cai, S.~Capozziello, M.~De~Laurentis, and E.~N. Saridakis, ``{f(T)
  teleparallel gravity and cosmology},''
  \href{http://dx.doi.org/10.1088/0034-4885/79/10/106901}{{\em Rept. Prog.
  Phys.} {\bf 79} (2016) no.~10, 106901},
\href{http://arxiv.org/abs/1511.07586}{{\tt arXiv:1511.07586 [gr-qc]}}.

\bibitem{Maluf:2013gaa}
J.~W. Maluf, ``{The teleparallel equivalent of general relativity},''
  \href{http://dx.doi.org/10.1002/andp.201200272}{{\em Annalen Phys.} {\bf 525}
  (2013)  339--357}, \href{http://arxiv.org/abs/1303.3897}{{\tt arXiv:1303.3897
  [gr-qc]}}.

\bibitem{aldrovandi1995introduction}
R.~Aldrovandi and J.~Pereira, {\em An Introduction to Geometrical Physics}.
\newblock World Scientific, 1995.
\newblock \url{https://books.google.com.mt/books?id=w8hBT4DV1vkC}.

\bibitem{Ferraro:2006jd}
R.~Ferraro and F.~Fiorini, ``{Modified teleparallel gravity: Inflation without
  inflaton},'' \href{http://dx.doi.org/10.1103/PhysRevD.75.084031}{{\em Phys.
  Rev. D} {\bf 75} (2007)  084031},
  \href{http://arxiv.org/abs/gr-qc/0610067}{{\tt arXiv:gr-qc/0610067}}.

\bibitem{Ferraro:2008ey}
R.~Ferraro and F.~Fiorini, ``{On Born-Infeld Gravity in Weitzenbock
  spacetime},'' \href{http://dx.doi.org/10.1103/PhysRevD.78.124019}{{\em Phys.
  Rev. D} {\bf 78} (2008)  124019}, \href{http://arxiv.org/abs/0812.1981}{{\tt
  arXiv:0812.1981 [gr-qc]}}.

\bibitem{Bengochea:2008gz}
G.~R. Bengochea and R.~Ferraro, ``{Dark torsion as the cosmic speed-up},''
  \href{http://dx.doi.org/10.1103/PhysRevD.79.124019}{{\em Phys. Rev. D} {\bf
  79} (2009)  124019}, \href{http://arxiv.org/abs/0812.1205}{{\tt
  arXiv:0812.1205 [astro-ph]}}.

\bibitem{Linder:2010py}
E.~V. Linder, ``{Einstein's Other Gravity and the Acceleration of the
  Universe},'' \href{http://dx.doi.org/10.1103/PhysRevD.81.127301}{{\em Phys.
  Rev. D} {\bf 81} (2010)  127301}, \href{http://arxiv.org/abs/1005.3039}{{\tt
  arXiv:1005.3039 [astro-ph.CO]}}. [Erratum: Phys.Rev.D 82, 109902 (2010)].

\bibitem{Chen:2010va}
S.-H. Chen, J.~B. Dent, S.~Dutta, and E.~N. Saridakis, ``{Cosmological
  perturbations in f(T) gravity},''
  \href{http://dx.doi.org/10.1103/PhysRevD.83.023508}{{\em Phys. Rev. D} {\bf
  83} (2011)  023508}, \href{http://arxiv.org/abs/1008.1250}{{\tt
  arXiv:1008.1250 [astro-ph.CO]}}.

\bibitem{Bahamonde:2019zea}
S.~Bahamonde, K.~Flathmann, and C.~Pfeifer, ``{Photon sphere and perihelion
  shift in weak $f(T)$ gravity},''
  \href{http://dx.doi.org/10.1103/PhysRevD.100.084064}{{\em Phys. Rev. D} {\bf
  100} (2019) no.~8, 084064}, \href{http://arxiv.org/abs/1907.10858}{{\tt
  arXiv:1907.10858 [gr-qc]}}.

\bibitem{Paliathanasis:2017htk}
A.~Paliathanasis, J.~Levi~Said, and J.~D. Barrow, ``{Stability of the Kasner
  Universe in f(T) Gravity},''
  \href{http://dx.doi.org/10.1103/PhysRevD.97.044008}{{\em Phys. Rev. D} {\bf
  97} (2018) no.~4, 044008}, \href{http://arxiv.org/abs/1709.03432}{{\tt
  arXiv:1709.03432 [gr-qc]}}.

\bibitem{Farrugia:2020fcu}
G.~Farrugia, J.~Levi~Said, and A.~Finch, ``{Gravitoelectromagnetism, Solar
  System Tests, and Weak-Field Solutions in $f (T,B)$ Gravity with
  Observational Constraints},''
  \href{http://dx.doi.org/10.3390/universe6020034}{{\em Universe} {\bf 6}
  (2020) no.~2, 34}, \href{http://arxiv.org/abs/2002.08183}{{\tt
  arXiv:2002.08183 [gr-qc]}}.

\bibitem{Bahamonde:2021srr}
S.~Bahamonde, A.~Golovnev, M.-J. Guzm\'an, J.~L. Said, and C.~Pfeifer, ``{Black
  holes in f(T,B) gravity: exact and perturbed solutions},''
  \href{http://dx.doi.org/10.1088/1475-7516/2022/01/037}{{\em JCAP} {\bf 01}
  (2022) no.~01, 037}, \href{http://arxiv.org/abs/2110.04087}{{\tt
  arXiv:2110.04087 [gr-qc]}}.

\bibitem{Bahamonde:2020bbc}
S.~Bahamonde, J.~Levi~Said, and M.~Zubair, ``{Solar system tests in modified
  teleparallel gravity},''
  \href{http://dx.doi.org/10.1088/1475-7516/2020/10/024}{{\em JCAP} {\bf 10}
  (2020)  024}, \href{http://arxiv.org/abs/2006.06750}{{\tt arXiv:2006.06750
  [gr-qc]}}.

\bibitem{Horndeski:1974wa}
G.~W. Horndeski, ``{Second-order scalar-tensor field equations in a
  four-dimensional space},'' \href{http://dx.doi.org/10.1007/BF01807638}{{\em
  Int. J. Theor. Phys.} {\bf 10} (1974)  363--384}.

\bibitem{TheLIGOScientific:2017qsa}
{\bf LIGO Scientific, Virgo} Collaboration, B.~P. Abbott {\em et al.},
  ``{GW170817: Observation of Gravitational Waves from a Binary Neutron Star
  Inspiral},'' \href{http://dx.doi.org/10.1103/PhysRevLett.119.161101}{{\em
  Phys. Rev. Lett.} {\bf 119} (2017) no.~16, 161101},
\href{http://arxiv.org/abs/1710.05832}{{\tt arXiv:1710.05832 [gr-qc]}}.

\bibitem{Goldstein:2017mmi}
A.~Goldstein {\em et al.}, ``{An Ordinary Short Gamma-Ray Burst with
  Extraordinary Implications: Fermi-GBM Detection of GRB 170817A},''
  \href{http://dx.doi.org/10.3847/2041-8213/aa8f41}{{\em Astrophys. J.} {\bf
  848} (2017) no.~2, L14},
\href{http://arxiv.org/abs/1710.05446}{{\tt arXiv:1710.05446 [astro-ph.HE]}}.

\bibitem{Ezquiaga:2017ekz}
J.~M. Ezquiaga and M.~Zumalac\'arregui, ``{Dark Energy After GW170817: Dead
  Ends and the Road Ahead},''
  \href{http://dx.doi.org/10.1103/PhysRevLett.119.251304}{{\em Phys. Rev.
  Lett.} {\bf 119} (2017) no.~25, 251304},
  \href{http://arxiv.org/abs/1710.05901}{{\tt arXiv:1710.05901 [astro-ph.CO]}}.

\bibitem{Bahamonde:2019shr}
S.~Bahamonde, K.~F. Dialektopoulos, and J.~Levi~Said, ``{Can Horndeski Theory
  be recast using Teleparallel Gravity?},''
  \href{http://dx.doi.org/10.1103/PhysRevD.100.064018}{{\em Phys. Rev. D} {\bf
  100} (2019) no.~6, 064018}, \href{http://arxiv.org/abs/1904.10791}{{\tt
  arXiv:1904.10791 [gr-qc]}}.

\bibitem{Gonzalez:2015sha}
P.~Gonzalez and Y.~Vasquez, ``{Teleparallel Equivalent of Lovelock Gravity},''
  \href{http://dx.doi.org/10.1103/PhysRevD.92.124023}{{\em Phys. Rev. D} {\bf
  92} (2015) no.~12, 124023}, \href{http://arxiv.org/abs/1508.01174}{{\tt
  arXiv:1508.01174 [hep-th]}}.

\bibitem{Bahamonde:2019ipm}
S.~Bahamonde, K.~F. Dialektopoulos, V.~Gakis, and J.~Levi~Said, ``{Reviving
  Horndeski theory using teleparallel gravity after GW170817},''
  \href{http://dx.doi.org/10.1103/PhysRevD.101.084060}{{\em Phys. Rev. D} {\bf
  101} (2020) no.~8, 084060}, \href{http://arxiv.org/abs/1907.10057}{{\tt
  arXiv:1907.10057 [gr-qc]}}.

\bibitem{Bahamonde:2021dqn}
S.~Bahamonde, M.~Caruana, K.~F. Dialektopoulos, V.~Gakis, M.~Hohmann,
  J.~Levi~Said, E.~N. Saridakis, and J.~Sultana, ``{Gravitational Wave
  Propagation and Polarizations in the Teleparallel analog of Horndeski
  Gravity},'' \href{http://arxiv.org/abs/2105.13243}{{\tt arXiv:2105.13243
  [gr-qc]}}.

\bibitem{Bahamonde:2020cfv}
S.~Bahamonde, K.~F. Dialektopoulos, M.~Hohmann, and J.~Levi~Said,
  ``{Post-Newtonian limit of Teleparallel Horndeski gravity},''
  \href{http://dx.doi.org/10.1088/1361-6382/abc441}{{\em Class. Quant. Grav.}
  {\bf 38} (2020) no.~2, 025006}, \href{http://arxiv.org/abs/2003.11554}{{\tt
  arXiv:2003.11554 [gr-qc]}}.

\bibitem{Dialektopoulos:2021ryi}
K.~F. Dialektopoulos, J.~L. Said, and Z.~Oikonomopoulou, ``{Classification of
  teleparallel Horndeski cosmology via Noether symmetries},''
  \href{http://dx.doi.org/10.1140/epjc/s10052-022-10201-7}{{\em Eur. Phys. J.
  C} {\bf 82} (2022) no.~3, 259}, \href{http://arxiv.org/abs/2112.15045}{{\tt
  arXiv:2112.15045 [gr-qc]}}.

\bibitem{Bernardo:2021bsg}
R.~C. Bernardo, J.~L. Said, M.~Caruana, and S.~Appleby, ``{Well-Tempered
  Minkowski Solutions in Teleparallel Horndeski Theory},''
  \href{http://arxiv.org/abs/2108.02500}{{\tt arXiv:2108.02500 [gr-qc]}}.

\bibitem{Bernardo:2021izq}
R.~C. Bernardo, J.~L. Said, M.~Caruana, and S.~Appleby, ``{Well-Tempered
  Teleparallel Horndeski Cosmology: A Teleparallel Variation to the
  Cosmological Constant Problem},'' \href{http://arxiv.org/abs/2107.08762}{{\tt
  arXiv:2107.08762 [gr-qc]}}.

\bibitem{Appleby:2018yci}
S.~Appleby and E.~V. Linder, ``{The Well-Tempered Cosmological Constant},''
  \href{http://dx.doi.org/10.1088/1475-7516/2018/07/034}{{\em JCAP} {\bf 07}
  (2018)  034}, \href{http://arxiv.org/abs/1805.00470}{{\tt arXiv:1805.00470
  [gr-qc]}}.

\bibitem{Capozziello:2023foy}
S.~Capozziello, M.~Caruana, J.~Levi~Said, and J.~Sultana, ``{Ghost and
  Laplacian Instabilities in the Teleparallel Horndeski Gravity},''
  \href{http://arxiv.org/abs/2301.04457}{{\tt arXiv:2301.04457 [gr-qc]}}.

\bibitem{misner1973gravitation}
C.~Misner, K.~Thorne, and J.~Wheeler, {\em Gravitation}.
\newblock No.~pt. 3 in Gravitation. W. H. Freeman, 1973.
\newblock \url{https://books.google.com.mt/books?id=w4Gigq3tY1kC}.

\bibitem{Weitzenbock1923}
R.~Weitzenb\"{o}ck, {\em `Invariantentheorie'}.
\newblock Noordhoff, Gronningen, 1923.

\bibitem{ortin2004gravity}
T.~Ort{\'\i}n, {\em Gravity and Strings}.
\newblock Cambridge Monographs on Mathematical Physics. Cambridge University
  Press, 2004.
\newblock \url{https://books.google.com.mt/books?id=sRlHoXdAVNwC}.

\bibitem{Krssak:2015oua}
M.~Krššák and E.~N. Saridakis, ``{The covariant formulation of f(T)
  gravity},'' \href{http://dx.doi.org/10.1088/0264-9381/33/11/115009}{{\em
  Class. Quant. Grav.} {\bf 33} (2016) no.~11, 115009},
\href{http://arxiv.org/abs/1510.08432}{{\tt arXiv:1510.08432 [gr-qc]}}.

\bibitem{PhysRevD.19.3524}
K.~Hayashi and T.~Shirafuji, ``New general relativity,''
  \href{http://dx.doi.org/10.1103/PhysRevD.19.3524}{{\em Phys. Rev. D} {\bf 19}
  (1979)  3524--3553}. \url{https://link.aps.org/doi/10.1103/PhysRevD.19.3524}.

\bibitem{Bahamonde:2017wwk}
S.~Bahamonde, C.~G. B{\"{o}}hmer, and M.~Kr{\v{s}}{\v{s}}{\'{a}}k, ``{New
  classes of modified teleparallel gravity models},''
  \href{http://dx.doi.org/10.1016/j.physletb.2017.10.026}{{\em Phys. Lett. B}
  {\bf 775} (2017)  37--43},
\href{http://arxiv.org/abs/1706.04920}{{\tt arXiv:1706.04920 [gr-qc]}}.

\bibitem{Bahamonde:2015zma}
S.~Bahamonde, C.~G. Böhmer, and M.~Wright, ``{Modified teleparallel theories
  of gravity},'' \href{http://dx.doi.org/10.1103/PhysRevD.92.104042}{{\em Phys.
  Rev.} {\bf D92} (2015) no.~10, 104042},
\href{http://arxiv.org/abs/1508.05120}{{\tt arXiv:1508.05120 [gr-qc]}}.

\bibitem{Finch:2018gkh}
A.~Finch and J.~L. Said, ``{Galactic Rotation Dynamics in f(T) gravity},''
  \href{http://dx.doi.org/10.1140/epjc/s10052-018-6028-1}{{\em Eur. Phys. J. C}
  {\bf 78} (2018) no.~7, 560}, \href{http://arxiv.org/abs/1806.09677}{{\tt
  arXiv:1806.09677 [astro-ph.GA]}}.

\bibitem{Farrugia:2016qqe}
G.~Farrugia and J.~Levi~Said, ``{Stability of the flat FLRW metric in $f(T)$
  gravity},'' \href{http://dx.doi.org/10.1103/PhysRevD.94.124054}{{\em Phys.
  Rev. D} {\bf 94} (2016) no.~12, 124054},
  \href{http://arxiv.org/abs/1701.00134}{{\tt arXiv:1701.00134 [gr-qc]}}.

\bibitem{Farrugia:2016xcw}
G.~Farrugia, J.~Levi~Said, and M.~L. Ruggiero, ``{Solar System tests in f(T)
  gravity},'' \href{http://dx.doi.org/10.1103/PhysRevD.93.104034}{{\em Phys.
  Rev. D} {\bf 93} (2016) no.~10, 104034},
  \href{http://arxiv.org/abs/1605.07614}{{\tt arXiv:1605.07614 [gr-qc]}}.

\bibitem{Iorio:2012cm}
L.~Iorio and E.~N. Saridakis, ``{Solar system constraints on f(T) gravity},''
  \href{http://dx.doi.org/10.1111/j.1365-2966.2012.21995.x}{{\em Mon. Not. Roy.
  Astron. Soc.} {\bf 427} (2012)  1555},
  \href{http://arxiv.org/abs/1203.5781}{{\tt arXiv:1203.5781 [gr-qc]}}.

\bibitem{Deng:2018ncg}
X.-M. Deng, ``{Probing f(T) gravity with gravitational time advancement},''
  \href{http://dx.doi.org/10.1088/1361-6382/aad391}{{\em Class. Quant. Grav.}
  {\bf 35} (2018) no.~17, 175013}.

\bibitem{BeltranJimenez:2020sih}
J.~Beltr{\'{a}}n~Jim{\'{e}}nez, L.~Heisenberg, and T.~Koivisto, ``{The coupling
  of matter and spacetime geometry},''
  \href{http://dx.doi.org/10.1088/1361-6382/aba31b}{{\em Class. Quant. Grav.}
  {\bf 37} (2020) no.~19, 195013},
\href{http://arxiv.org/abs/2004.04606}{{\tt arXiv:2004.04606 [hep-th]}}.

\bibitem{Kobayashi:2011nu}
T.~Kobayashi, M.~Yamaguchi, and J.~Yokoyama, ``{Generalized G-inflation:
  Inflation with the most general second-order field equations},''
  \href{http://dx.doi.org/10.1143/PTP.126.511}{{\em Prog. Theor. Phys.} {\bf
  126} (2011)  511--529}, \href{http://arxiv.org/abs/1105.5723}{{\tt
  arXiv:1105.5723 [hep-th]}}.

\bibitem{Bellini:2014fua}
E.~Bellini and I.~Sawicki, ``{Maximal freedom at minimum cost: linear
  large-scale structure in general modifications of gravity},''
  \href{http://dx.doi.org/10.1088/1475-7516/2014/07/050}{{\em JCAP} {\bf 07}
  (2014)  050}, \href{http://arxiv.org/abs/1404.3713}{{\tt arXiv:1404.3713
  [astro-ph.CO]}}.

\bibitem{Dialektopoulos:2023dhb}
K.~F. Dialektopoulos, P.~Mukherjee, J.~Levi~Said, and J.~Mifsud, ``{Neural
  network reconstruction of cosmology using the Pantheon Compilation},''
  \href{http://arxiv.org/abs/2305.15499}{{\tt arXiv:2305.15499 [gr-qc]}}.

\bibitem{Dialektopoulos:2023jam}
K.~F. Dialektopoulos, P.~Mukherjee, J.~Levi~Said, and J.~Mifsud, ``{Neural
  network reconstruction of scalar-tensor cosmology},''
  \href{http://arxiv.org/abs/2305.15500}{{\tt arXiv:2305.15500 [gr-qc]}}.

\bibitem{Briffa:2023ern}
R.~Briffa, C.~Escamilla-Rivera, J.~Levi~Said, and J.~Mifsud, ``{Constraints on
  f(T) cosmology with Pantheon+},''
  \href{http://dx.doi.org/10.1093/mnras/stad1384}{{\em Mon. Not. Roy. Astron.
  Soc.} {\bf 522} (2023) no.~4, 6024--6034},
  \href{http://arxiv.org/abs/2303.13840}{{\tt arXiv:2303.13840 [gr-qc]}}.

\bibitem{Dialektopoulos:2021wde}
K.~Dialektopoulos, J.~L. Said, J.~Mifsud, J.~Sultana, and K.~Z. Adami,
  ``{Neural network reconstruction of late-time cosmology and null tests},''
  \href{http://dx.doi.org/10.1088/1475-7516/2022/02/023}{{\em JCAP} {\bf 02}
  (2022) no.~02, 023}, \href{http://arxiv.org/abs/2111.11462}{{\tt
  arXiv:2111.11462 [astro-ph.CO]}}.

\bibitem{LeviSaid:2021yat}
J.~Levi~Said, J.~Mifsud, J.~Sultana, and K.~Z. Adami, ``{Reconstructing
  teleparallel gravity with cosmic structure growth and expansion rate data},''
  \href{http://dx.doi.org/10.1088/1475-7516/2021/06/015}{{\em JCAP} {\bf 06}
  (2021)  015}, \href{http://arxiv.org/abs/2103.05021}{{\tt arXiv:2103.05021
  [astro-ph.CO]}}.

\bibitem{Hwang:2022hla}
S.-g. Hwang, B.~L'Huillier, R.~E. Keeley, M.~J. Jee, and A.~Shafieloo, ``{How
  to use GP: effects of the mean function and hyperparameter selection on
  Gaussian process regression},''
  \href{http://dx.doi.org/10.1088/1475-7516/2023/02/014}{{\em JCAP} {\bf 02}
  (2023)  014}, \href{http://arxiv.org/abs/2206.15081}{{\tt arXiv:2206.15081
  [astro-ph.CO]}}.

\bibitem{Planck:2018vyg}
{\bf Planck} Collaboration, N.~Aghanim {\em et al.}, ``{Planck 2018 results.
  VI. Cosmological parameters},''
  \href{http://dx.doi.org/10.1051/0004-6361/201833910}{{\em Astron. Astrophys.}
  {\bf 641} (2020)  A6}, \href{http://arxiv.org/abs/1807.06209}{{\tt
  arXiv:1807.06209 [astro-ph.CO]}}. [Erratum: Astron.Astrophys. 652, C4
  (2021)].

\bibitem{Saltas:2014dha}
I.~D. Saltas, I.~Sawicki, L.~Amendola, and M.~Kunz, ``{Anisotropic Stress as a
  Signature of Nonstandard Propagation of Gravitational Waves},''
  \href{http://dx.doi.org/10.1103/PhysRevLett.113.191101}{{\em Phys. Rev.
  Lett.} {\bf 113} (2014) no.~19, 191101},
  \href{http://arxiv.org/abs/1406.7139}{{\tt arXiv:1406.7139 [astro-ph.CO]}}.

\bibitem{Riazuelo:2000fc}
A.~Riazuelo and J.-P. Uzan, ``{Quintessence and gravitational waves},''
  \href{http://dx.doi.org/10.1103/PhysRevD.62.083506}{{\em Phys. Rev. D} {\bf
  62} (2000)  083506}, \href{http://arxiv.org/abs/astro-ph/0004156}{{\tt
  arXiv:astro-ph/0004156}}.

\bibitem{Bellini}
E.~{Bellini} and I.~{Sawicki}, ``{Maximal freedom at minimum cost: linear
  large-scale structure in general modifications of gravity},''
  \href{http://dx.doi.org/10.1088/1475-7516/2014/07/050}{{\em "JCAP"} {\bf
  2014} (2014) no.~7, 050}, \href{http://arxiv.org/abs/1404.3713}{{\tt
  arXiv:1404.3713 [astro-ph.CO]}}.

\end{thebibliography}\endgroup

\appendix

\section{Teleparallel scalars in the cosmological setting}
Here, all independent scalars in BDLS theory are presented explicitly for tensor, vector and scalar perturbations separately. The ones that are not mentioned are vanishing.

The non-trivial scalars of tensor perturbation take the following form
\begin{gather}
    I_2=\frac{3\dot{a}\dot{\phi}}{a},\\
    J_5=\frac{\dot{\phi}^2}{8}(\,\dot{h}_+^2+\dot{h}_{\times}^2\,),\\    J_8=\frac{\dot{\phi}^2}{2}(\,\dot{h}_+^2+\dot{h}_{\times}^2\,).
\end{gather}
The respective ones in vector perturbations are
\begin{gather}
    I_{2}=\frac{3 \dot{a} \dot{\phi}}{a}+\frac{1}{a} \Big\{\dot{\phi} \left[-3 \dot{a}\left(2 \bm{b} \bm{\beta}+\bm{\beta}^2 \right)+ a \big(\bm{b} \dot{\bm{\beta}}-2 (\bm{\nabla}\times \bm{h}) (\bm{\nabla} \times\dot{\bm{h}})\big)+ (\bm{\nabla}\times \bm{b}) (\bm{\nabla}\times \bm{h}) \right] +\dot{\phi} (\bm{b}+\bm{\beta}) \left(\bm{\nabla}^2 \bm{h}-a \dot{\bm{\beta}}\right)\Big\}, 
    \\
    J_{3}=\frac{1}{6a^2} \Big\{\dot{\phi}^2 \left[2 a^2 \dot{\bm{\beta}}^2- (\bm{\nabla}^2 \bm{h})^2-a\dot{\bm{\beta}} \bm{\nabla}^2 \bm{h} \right]\Big\},
    \\
    J_{5}=\frac{1}{4a^2} \Bigg\{ \dot{\phi}^2 \left[\bm{\nabla}^2\times(-a \dot{\bm{h}}+\bm{b}+\bm{\beta})\right]^2+\dot{\phi}^2 \left(a \bm{\nabla}\times\dot{\bm{h}}+\bm{\nabla}\times \bm{\beta} \right)^2-\frac{5 \dot{\phi}^2 \left(2 a \dot{\bm{\beta}}+\Delta \bm{h}\right)^2}{9} \Bigg\},
    \\
    J_{6}=\frac{1}{36a^2} \Bigg[\dot{\phi}^4 \left(2 a \dot{\bm{\beta}}+\bm{\nabla}^2 \bm{h}\right)^2\Bigg],
    \\
    J_{8}=\frac{1}{2a^2} \Bigg[\dot{\phi}^2 \left(2 a \bm{\nabla}\times\dot{\bm{h}}-\bm{\nabla}\times \bm{b} \right)^2-\frac{\dot{\phi}^2 \left(2 a \dot{\bm{\beta}}+\Delta \bm{h}\right)^2}{9}\Bigg],
    \\
    J_{10}=\frac{\dot{\phi}^2}{6 a^2} \bigg[ (\bm{\nabla}\times \bm{b}) (\bm{\nabla}\times\bm{\beta})+(\bm{\nabla}\times \bm{b})^2-2(\bm{\nabla}\times\bm{\beta})^2 \bigg],
\end{gather}
and the ones related to scalar perturbations are%
 \begin{gather}
    I_2=\frac{3\dot{a} \dot{\phi}}{a}-\frac{\dot{\phi}}{a}\,\left[6 \alpha \dot{a}+a(\bm{\nabla}^2{\beta}-3\dot{\xi})-9\alpha^2\dot{a}-2 a\alpha(\bm{\nabla}^2{\beta}-3\dot{\xi})+3 a\bm{\nabla}{\beta}\bm{\nabla}{\xi}\right],\\[7pt]
    J_3=\frac{\dot{\phi}^2}{3a^2}\,\left[\, (\bm{\nabla}\alpha)^2+\bm{\nabla}\zeta\,\bm{\nabla}\alpha-2\,(\bm{\nabla}\zeta)^2\,\right],\\[7pt]
    J_5=\frac{\dot{\phi}^2}{18a^2}\,\left[\,3a^2\,(\bm{\nabla}^2\beta)^2-10\,(\bm{\nabla}\zeta-\bm{\nabla}\alpha)^2\,\right],\\[7pt]
    J_6=\frac{\dot{\phi}^4}{9a^2}\,\left(\bm{\nabla}\alpha-\bm{\nabla}\zeta\,\right)^2,\\[7pt]
    J_8=\frac{2\dot{\phi}^2}{9a^2}\,\left[\,3a^2\,(\bm{\nabla}^2\beta)^2-(\bm{\nabla}\zeta-\bm{\nabla}\alpha)^2\,\right] \,.
 \end{gather}

\end{document}